\documentclass{article}
\usepackage[english]{babel}
\usepackage[utf8]{inputenc}
\usepackage{csquotes}
\usepackage{graphicx}
\usepackage{float}
\usepackage{geometry}
\usepackage{parskip}
\usepackage{placeins}
\usepackage{braket}
\usepackage{amsmath}
\usepackage{amsfonts}
\usepackage{bbm}
\usepackage{bm}
\usepackage{xcolor}
\usepackage{tabularx}
\usepackage[hidelinks]{hyperref}
\usepackage{cleveref}
\usepackage{numprint}
\usepackage{booktabs}       
\usepackage{makecell}
\usepackage[backend=biber, sorting=none, style=nature, citestyle=numeric-comp, maxcitenames=1]{biblatex}
\usepackage{siunitx}
\usepackage{chemformula}
\usepackage{multicol}
\usepackage{subcaption}
\usepackage{wrapfig}
\usepackage{overpic}
\usepackage{authblk}
\usepackage{titlesec}  
\Crefname{appendix}{App.}{Apps.}
\Crefname{figure}{Fig.}{Figs.}
\Crefname{table}{Tab.}{Tabs.}
\Crefname{equation}{Eq.}{Eqs.}
\Crefname{section}{Sec.}{Secs.}
\crefname{appendix}{App.}{Apps.}
\crefname{section}{Sec.}{Secs.}
\crefname{figure}{Fig.}{Figs.}
\crefname{table}{Tab.}{Tabs.}
\crefname{equation}{Eq.}{Eqs.}

\npthousandsep{,}

\usepackage{amsmath,amsfonts,bm,bbm,amssymb}

\newcommand{\citet}[1]{\citeauthor{#1} (\citeyear{#1})}
\newcommand{\citep}[1]{\cite{#1}}

\def\eqref#1{equation~\ref{#1}}

\def\1{\bm{1}}

\def\vb{{\bm{b}}}

\def\ve{{\bm{e}}}

\def\vg{{\bm{g}}}
\def\vh{{\bm{h}}}

\def\vk{{\bm{k}}}

\def\vm{{\bm{m}}}

\def\vq{{\bm{q}}}
\def\vr{{\bm{r}}}

\def\vv{{\bm{v}}}
\def\vw{{\bm{w}}}
\def\vx{{\bm{x}}}

\def\mA{{\bm{A}}}

\def\mH{{\bm{H}}}

\def\mU{{\bm{U}}}
\def\mV{{\bm{V}}}
\def\mW{{\bm{W}}}

\DeclareMathAlphabet{\mathsfit}{\encodingdefault}{\sfdefault}{m}{sl}
\SetMathAlphabet{\mathsfit}{bold}{\encodingdefault}{\sfdefault}{bx}{n}

\newcommand{\E}{\mathbb{E}}

\newcommand{\R}{\mathbb{R}}

\newcommand{\softmax}{\mathrm{softmax}}

\DeclareMathOperator{\Tr}{Tr}

\def\wf{{\Psi}}
\def\orbs{{\Phi}}
\def\emb{{\vh}}

\def\wfp{{\rho_\wf}}
\def\rvec{{\vr}}
\def\r{{r}} 
\def\N{{\mathcal{N}}}
\def\cutoff{{c}}
\def\ncutoff{{c_\text{nuc}}}
\newcommand{\Ewf}[1][\eposs]{\E_{\wfp}}
\newcommand{\lwf}[1][(\eposs)]{\ln\vert\wf#1\vert}
\newcommand{\glwf}[1][(\eposs)]{\nabla_\theta\lwf[#1]}

\def\H{{\hat{H}}}
\def\El{{E_L}}
\def\qgt{\mathcal{O}}

\def\mcqgt{\bm{\bar{\qgt}}}

\def\epos{{\vec{r}}}
\def\npos{{\vec{R}}}
\def\eposs{{\vec{\mathbf{r}}}}
\def\nposs{{\vec{\mathbf{R}}}}
\def\charge{{Z}}
\def\charges{{\mathbf{Z}}}

\def\Nreg{{N_\text{reg}}}
\def\dreg{{d_\text{reg}}}
\def\Nquad{{N_\text{quad}}}

\def\Nnuc{{N_\mathrm{n}}}

\def\Nwalker{{N_\mathrm{walker}}}

\def\Nup{{N_\uparrow}}
\def\Ndown{{N_\downarrow}}

\def\Nenv{{N_\mathrm{env}}}

\def\Jastrow{{\mathcal{J}}}

\def\MLP{\text{MLP}}

\DeclareSIUnit\bohr{\ensuremath{a_0}}
\DeclareSIUnit\hartree{\text{\ensuremath{E_\textup{h}}}}
\DeclareSIUnit\calorine{cal}
\DeclareSIUnit\kcal{\kilo\calorine}

\renewcommand{\vec}[1]{{\bm{#1}}}

\newcommand{\nel}{{n_{\mathrm{el}}}}
\newcommand{\ndim}{{n_\text{dim}}}
\newcommand{\nnb}{{n_{\mathrm{nb}}}}

\newcommand{\nbset}{\mathcal{N}}
\newcommand{\updset}{\mathcal{U}}

\newcommand{\ndet}{N_{\mathrm{det}}}

\renewcommand{\O}{\mathcal{O}}

\newcommand{\Twf}{T_\text{full wf}}
\newcommand{\Tupd}{T_\text{upd}}
\newcommand{\Tsamp}{T_\text{sampling}}
\newcommand{\Tkin}{T_\text{kinetic}}
\newcommand{\Tecp}{T_\text{ECP}}
\newcommand{\Tspin}{T_\text{spin}}
\newcommand{\Ttot}{T_\text{tot}}
\newcommand{\globalstep}{\sigma_\text{g}}
\newcommand{\nucemb}{\hat{\vh}^\text{nuc}}

\DeclareSIUnit{\angstrom}{\text{\normalfont\accent23A}}
\newcommand{\bohr}{$a_0$}
\DeclareSIUnit\mHa{mE_h}
\AtEveryBibitem{%
  \ifentrytype{article}{%
    \clearfield{urlyear}%
    \clearfield{url}%
    \clearfield{eprint}%
    \clearfield{issn}%
  }{}%
  \ifentrytype{book}{%
    \clearfield{urlyear}%
    \clearfield{url}%
    \clearfield{eprint}%
    \clearfield{issn}%
  }{}%
}
\DeclareFieldFormat{eprint:arxiv}{%
  arXiv:\href{https://arxiv.org/abs/#1}{#1}} 
\DeclareFieldFormat{eprint}{%
  \mkbibbrackets{arXiv:\href{https://arxiv.org/abs/#1}{#1}}} 

\DeclareBibliographyDriver{misc}{%
  \usebibmacro{bibindex}%
  \usebibmacro{begentry}%
  \printnames{author}%
  \setunit{\labelnamepunct}%
  \printfield{title}%
  \newunit\newblock%
  \printfield{year}%
  \newunit\newblock%
  \printfield{month}%
  \newunit\newblock%
  \iftoggle{bbx:eprint}
    {\printfield[eprint]{eprint}}
    {}%
  \newunit\newblock%
  \usebibmacro{finentry}
}

\newenvironment{Figure}
  {\par\medskip\noindent\minipage{\linewidth}}
  {\endminipage\par\medskip}

\title{Accurate Ab-initio Neural-network Solutions to\\Large-Scale Electronic Structure Problems}
\author[1*]{Michael Scherbela}
\author[2*]{Nicholas Gao}
\author[1,3]{Philipp Grohs}
\author[2]{Stephan Günnemann}
\affil[1]{Faculty of Mathematics, University of Vienna, Oskar-Morgenstern-Platz 1, A-1090 Vienna, Austria}
\affil[2]{Department of Computer Science \& Munich Data Science Institute, 
Technical University of Munich}
\affil[3]{Johann Radon Institute for Computational and Applied Mathematics, Austrian Academy of Sciences\\
Altenbergerstrasse 69, 4040 Linz, Austria}
\affil[*]{Equal contribution, order determined by coin flip}
\date{}                     
\begin{document}

\maketitle
\begin{abstract}
\noindent
    We present finite-range embeddings (FiRE), a novel wave function ansatz for accurate large-scale \emph{ab-initio} electronic structure calculations.
    Compared to contemporary neural-network wave functions, FiRE reduces the asymptotic complexity of neural-network variational Monte Carlo (NN-VMC) by $\sim\nel$, the number of electrons.
    By restricting electron-electron interactions within the neural network, FiRE accelerates all key operations -- sampling, pseudopotentials, and Laplacian computations -- resulting in a real-world $10\times$ acceleration in now-feasible 180-electron calculations. 
    We validate our method's accuracy on various challenging systems, including biochemical compounds, conjugated hydrocarbons, and organometallic compounds.
    On these systems, FiRE's energies are consistently within chemical accuracy of the most reliable data, including experiments, even in cases where high-accuracy methods such as CCSD(T), AFQMC, or contemporary NN-VMC fall short.
    With these improvements in both runtime and accuracy, FiRE represents a new `gold-standard' method for fast and accurate large-scale ab-initio calculations, potentially enabling new computational studies in fields like quantum chemistry, solid-state physics, and material design.
\end{abstract}

\begin{multicols}{2}
\section{Introduction}
Solving the electronic Schrödinger equation unlocks the computational analysis of molecular and material properties and structures.
Unfortunately, its solution, the ground-state electronic wave function, is only known analytically for the simplest of systems.
Consequently, approximations trade off computational efficiency and accuracy on various scales depending on the problem, its properties, and the computational budget.
Some methods, such as Density Functional Theory (DFT), scale favorably with system size but fail to predict experiments for strongly correlated systems.
Other methods, such as Coupled Cluster, often correctly predict experiments, but their computational cost increases dramatically with the number of electrons $\nel$, e.g., $\O(\nel^7)$ for CCSD(T).
Furthermore, applying these highly accurate methods frequently requires expert knowledge in choosing basis sets, initialization, active spaces, and optimization parameters, even for small systems.

In theory, Variational Monte Carlo (VMC) promises both a favorable runtime by scaling only $\O(\nel^3)$ per step in the number of electrons $\nel$, and being easy to apply, as it directly parametrizes the real-space electron wave function $\wf:\R^{\nel\times 3}\to\R$~\citep{foulkesQuantumMonteCarlo2001}.
However, conventional VMC has long been touted in practice as a low-accuracy method that may only be used as initial guesses for accurate simulations like diffusion Monte Carlo~\citep{umrigarDiffusionMonteCarlo1993}.
This fundamentally changed with the recent advent of neural-network VMC (NN-VMC), which use a neural-network ansatz for the wave function. Due to the superior expressive power of neural networks compared to classical ansatze, NN-VMC frequently achieves the to-date most accurate energies for small molecules.  However, this gain in accuracy comes at the price of an increased cost of $\O(\nel^4)$ per step, which severely limits the system sizes for which the method is computationally tractable~\citep{pfauInitioSolutionManyelectron2020}.
The slowdown arises because contemporary neural wave functions do not support two critical operations that are needed in VMC: (1) efficient Laplacian calculations, which are necessary for energy evaluation, and (2) wave function updates if few electrons are moved, which are crucial for sampling and pseudopotentials.
Thus, there exists a clear gap between both flavors as conventional VMC is scalable but inaccurate, and NN-VMC is slow but accurate. The purpose of this paper is to significantly narrow this gap, as we will now describe.

In conventional VMC, one typically chooses a Slater-Jastrow wave function
\begin{align}
    \wf(\rvec) = \Jastrow(\rvec)\det\left[\Phi(\rvec)\right] \label{eq:slater_jastrow}
\end{align}
where a symmetric Jastrow factor $\Jastrow:\R^{\nel\times 3}\to\R$ is multiplied with a Slater determinant $\det[\Phi(\rvec)]$, which enforces fermionic antisymmetry.
For readability, we have omitted spin and limited the model to a single determinant.
In the absence of a backflow ~\citep{feynmanEnergySpectrumExcitations1956,luoBackflowTransformationsNeural2019}, the orbital matrix $\Phi(\rvec)  = [\Phi_{i l}(\epos)]_{i,l=1}^{\nel}\in\R^{\nel \times \nel}$ consists of single-electron orbitals $\phi_l:\R^3\to\R$:
\begin{align}
    \Phi_{il}(\rvec) = \phi_l(\rvec_i).\label{eq:single_orbitals}
\end{align}
While the orbitals being single-electron functions enables efficient single-electron updates and Laplacian calculations of the wave function in $\O(\nel^2)$ and $\O(\nel^3)$, respectively~\citep{fahyVariationalQuantumMonte1990}, their constrained functional form prohibits the accurate representation of strongly correlated systems~\citep{mottaSolutionManyElectronProblem2017}.
\begin{figure*}[!ht]
    \centering
    \includegraphics[width=\textwidth]{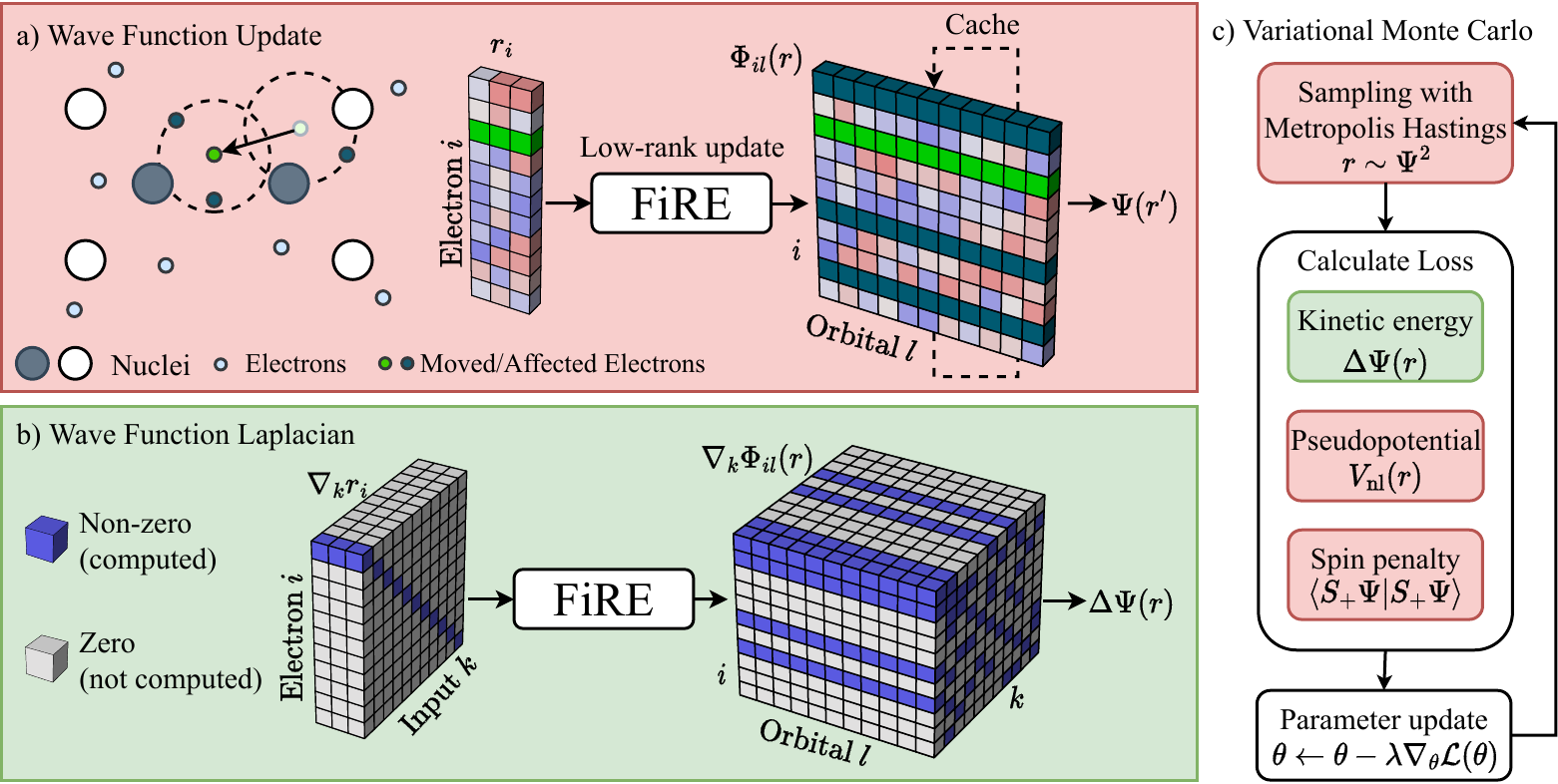}
    \caption{\textbf{Conceptual overview}: \textbf{a)} For each-single electron move, e.g., during sampling or pseudopotentials, we only update the orbitals of electrons within the cutoff of its old and new positions. \textbf{b)} FiRE enables efficient Laplacian computations by exploiting the sparsity patterns within the Jacobian $\nabla \orbs(\rvec)$ to only compute non-zero entries. \textbf{c)} All components of VMC that are accelerated by FiRE.}
    \label{fig:overview}
\end{figure*}

In NN-VMC, one lifts this constraint by replacing the single-electron orbitals with many-electron neural networks $\vh_i:\R^{\nel\times 3}\to\R^\ndim$ multiplied with envelope functions $\varphi_l:\R^3\to\R$ to ensure the correct long-range behavior:
\begin{align}
    \Phi_{il}(\rvec) &= (\vh_i(\rvec)^T \vw_l)\varphi_l(\rvec_i), \label{eq:orbitals}\\
    \vh_i(\rvec) &= \vh\left(\rvec_i, \left\{\rvec_j\right\}_{j\neq i}\right).
\end{align}
The $i$th electron's embedding $\vh_i(\rvec)$ depends in this formulation on the position of all other electrons, indicated by the (multi-)set $\left\{\epos_j\right\}_{j\neq i}$.
By choosing permutation-equivariant architectures like graph neural networks~\citep{hermannDeepneuralnetworkSolutionElectronic2020,gaoGeneralizingNeuralWave2023}, or transformers~\citep{vonglehnSelfAttentionAnsatzAbinitio2023} for $\vh$, antisymmetry is preserved.
For several small molecules, NN-VMC achieves energy estimates outperforming conventional `gold-standard' methods such as Coupled Clusters (CCSD(T)) or Multireference Configuration Interaction (MRCI) \cite{pfauInitioSolutionManyelectron2020, gerardGoldstandardSolutionsSchrodinger2022}.
However, as the orbital matrix elements depend on all electrons, the Laplacian requires $\O(\nel^4)$ operations, and efficient low-rank updates are impossible.
Several fruitful improvements have reduced the cost of NN-VMC, accelerating each optimization step~\citep{liFermionicNeuralNetwork2022,liComputationalFrameworkNeural2024}, reducing the number of optimization steps required~\citep{rendeSimpleLinearAlgebra2023,goldshlagerKaczmarzinspiredApproachAccelerate2024a}, or amortizing the cost across several systems~\citep{scherbelaTransferableFermionicNeural2024,gaoNeuralPfaffiansSolving2024}.
Nevertheless, none of these change the overall computational complexity, and system sizes studied by NN-VMC are typically still limited to $\nel\approx 80$ in 10,000 GPU hours~\citep{vonglehnSelfAttentionAnsatzAbinitio2023}.

In this work, we connect the favorable computational scaling of conventional VMC and the accuracy of NN-VMC by introducing a novel neural wave function based on finite-range embeddings (FiRE).
FiRE reduces the computational scaling of NN-VMC by $\O(\nel)$ to $\O(\nel^3)$, yielding speed-ups of $\approx 10\times$ for relevant system sizes.
This enables the application to larger systems at lower runtimes while maintaining highly accurate relative energies, yielding a new `gold standard' for fast and accurate \emph{ab-initio} electronic structure calculations.

\begin{figure*}[!ht]
    \centering
    \includegraphics[width=\textwidth]{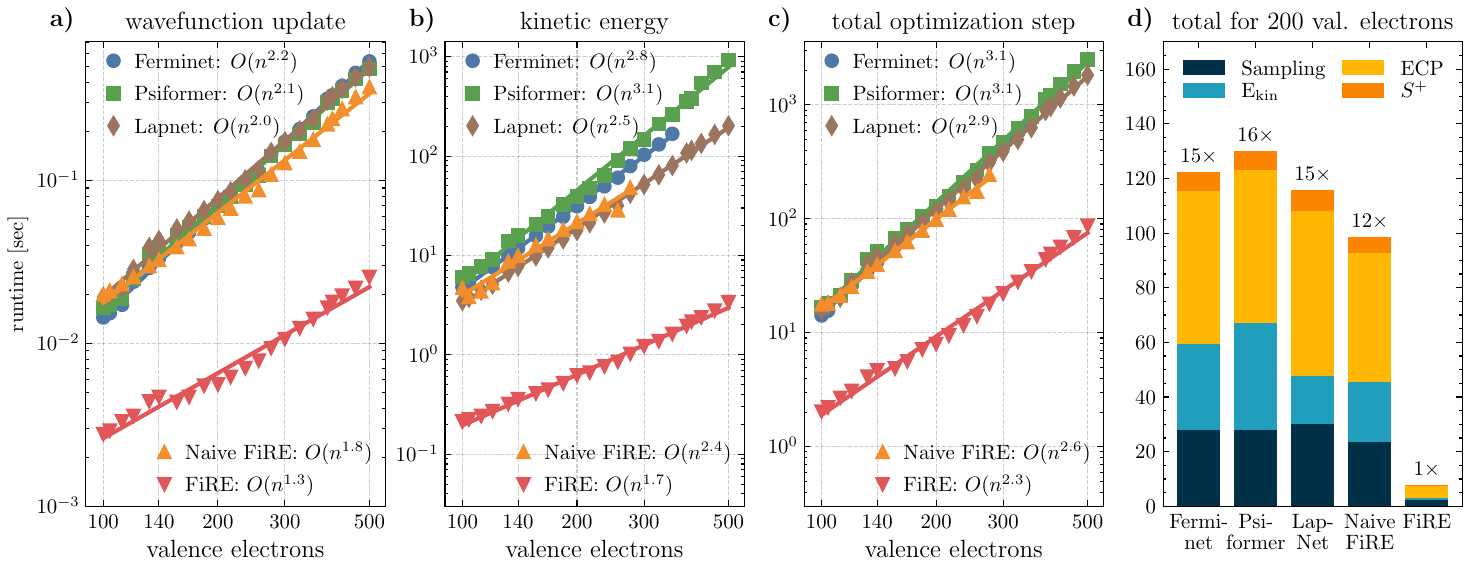}
    \caption{\textbf{Runtime for cumulene chains of varying length}. Runtimes for equivalent batch size of 4096 on a single A100 GPU. FiRE models use a cutoff $\cutoff=3a_0$ \textbf{a)} Time required to update the wave function $\Psi$ after single-electron move. \textbf{b)} Time required to compute the kinetic energy $\Delta \Psi$. \textbf{c)} Total time per optimization step. \textbf{d)} Breakdown of the runtime of a single optimization step for different architectures.}
    \label{fig:scaling} 
\end{figure*}

As electron interactions are most pronounced at short ranges, we focus the electron's embedding on its vicinity by limiting its dependence on the electrons within some cutoff $\cutoff$.
Thus, instead of the extrema of single-electron conventional VMC and all-electron NN-VMC, FiRE defines the $i$th electron's embedding, from which the orbital matrix $\Phi(\rvec)$ is derived (\cref{eq:orbitals}), via the neighborhood $\N_{\r_i}$ defined by the cutoff $c\in\R_+$:
\begin{align}
    \emb_i(\rvec) &= \emb(\rvec_i, \{\rvec_j\}_{j \in \N_{\r_i}}), \\
    \N_{\epos_i} &= \left\lbrace j\; \middle|\;  j \neq i \land \vert\rvec_i - \rvec_j\vert\leq \cutoff\right\rbrace.
\end{align}
This ansatz can trivially represent single-electron orbitals (\cref{eq:single_orbitals}), including arbitrarily delocalized orbitals, by letting $\cutoff=0$, i.e, $\N_{\epos_i}=\emptyset$.
Further, thanks to the dependence on the close-by electrons, FiRE is a superset of Slater-Jastrow wave functions as $\Jastrow$ may represent any many-body Jastrow factor depending on all electrons within the cutoff.
As $\cutoff \to \infty$, $\N_{\epos_i}\to\{\epos_j\}_{i\neq j}$ approaches the fully-connected limit and we recover contemporary neural wave functions (\cref{eq:orbitals}).
To capture electron correlation beyond the cutoff, we introduce a novel global attention-based Jastrow factor, detailed in \cref{sec:jastrow}.

We show in \cref{sec:improved_runtime} that FiRE speeds up all relevant aspects of NN-VMC by $\O(\nel)$, in particular sampling from the wave function, evaluating its energy and energy gradient, and evaluating non-local operators required for effective core potentials (ECPs), or spin-related quantities as visualized in \cref{fig:overview}.

In \cref{sec:accurate_energies}, we demonstrate the accuracy of our approach by applying it to various challenging systems, such as non-covalent interactions, large hydrocarbons, and organometallic compounds.
We find that even small cutoffs yield highly accurate wave functions, and compared to existing NN-VMC approaches, we obtain more accurate relative energies at a fraction of computational cost.
On several of these systems, we find FiRE to accurately reconstruct experimental results, even obtaining better predictions than `gold-standard' methods like CCSD(T) or AFQMC.

As we can now scale NN-VMC to unprecedented sizes, we analyze the convergence rates of NN-VMC both in system sizes and optimization steps in \cref{sec:convergence_rates}. Interestingly, we observe consistent convergence rates across different systems.

\section{Results}

\subsection{Improved runtime}
\label{sec:improved_runtime}
Computing the embeddings $\vh$ for fully connected architectures scales as $\O(\nel^2)$ due to the pairwise electron-electron interactions and is typically the computational bottleneck for medium-sized molecules.
However, evaluating the determinant, which scales $\O(\nel^3)$, determines the asymptotic scaling of the wave function.
Thus, replacing a fully connected embedding with our finite-range embedding (FiRE) does not change the asymptotic scaling and only provides modest speedups as shown in \cref{fig:scaling}d, where we compare this `Naive FiRE' against state-of-the-art neural wave functions.
The key advantage of FiRE is that its sparsity enables us to speed up two critical operations that determine the actual scaling of VMC: updating $\wf(\rvec)$ after moving a small number of electrons and computing the Laplacian $\Delta \wf$.

Several operations of a VMC optimization require wave function updates, i.e., evaluating $\wf(\rvec')$ when $\wf(\rvec)$ is known, and $\rvec'$ differs from $\rvec$ in only a few electrons' positions.
This occurs during Monte Carlo sampling, where new electron coordinates $\rvec'$ are proposed at each Markov Chain step via single-electron updates from $\rvec$.
Wave function updates are also necessary when evaluating non-local parts for effective core potentials (ECPs) and spin-operators, such as $S^2$~\citep{szaboImprovedPenaltybasedExcitedstate2024} or $S^+$~\citep{liSpinsymmetryenforcedSolutionManybody2024}.
In all three cases, a single optimization step typically requires $\O(\nel)$ updates, yielding the naive asymptotic per-step cost of $\O(\nel^4)$.
However, when using FiRE, we can exploit that moving a single electron affects only the embeddings and orbitals of electrons in its old and new neighborhood.
Instead of fully recomputing all orbitals, we only recompute the affected electrons (see \cref{fig:overview}a) and, instead of naively computing the determinant $\nel\times\nel$, we use low-rank updates scaling as $\O(\nel^2)$, as shown in \cref{sec:method_low_rank_update}.
These low-rank updates reduce the scaling of our updates by $\O(\nel)$.
\Cref{fig:scaling}a shows that we can obtain similar speedups in practice: While for previous neural wave functions computing a wavefunction update empirically scales between $\Tupd \sim \nel^{2.0}$ and $\Tupd \sim \nel^{2.2}$, FiRE only grows as $\Tupd \sim \nel^{1.3}$, achieving an approximate speedup proportional to $\nel$.

A similar improvement can be applied to the evaluation of the kinetic energy, which requires the Laplacian of $\wf$,
which in turn requires Jacobians of all intermediates of the neural network, including the orbitals $\Phi$.
In existing neural wave functions, every entry of $\Phi$ depends on every electron, and therefore the Jacobian $\nabla_\rvec \orbs$ is dense, containing $\O(\nel^3)$ entries.
Propagating this Jacobian forward requires $\O(\nel^4)$ operations as detailed in \cref{sec:Laplacian}.
In contrast, FiRE's Jacobian is sparse as depicted in \cref{fig:overview}b, yielding an $\O(\nel)$ speedup, which we again can see in empirical runtimes in \cref{fig:scaling}b.

\begin{figure*}[ht]
    \centering
    \includegraphics[width=\textwidth]{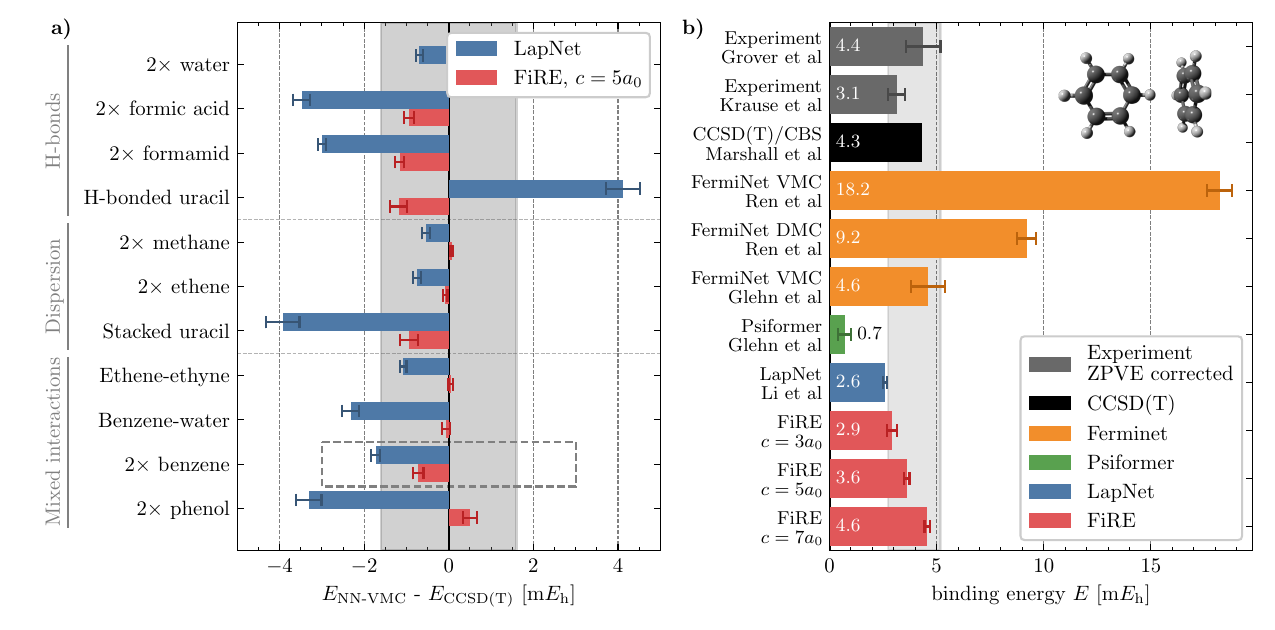}
    \vspace{-1em}
    \begin{minipage}[t]{\linewidth}
        \resizebox{1\linewidth}{!}{
        \begin{overpic}{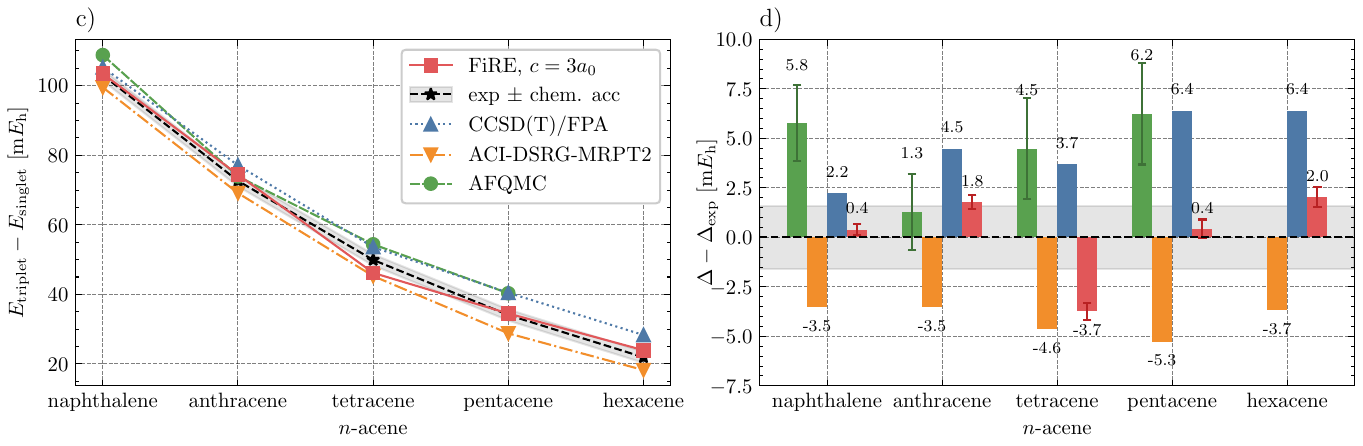}
        \put(5,5){\includegraphics[width=.25\linewidth]{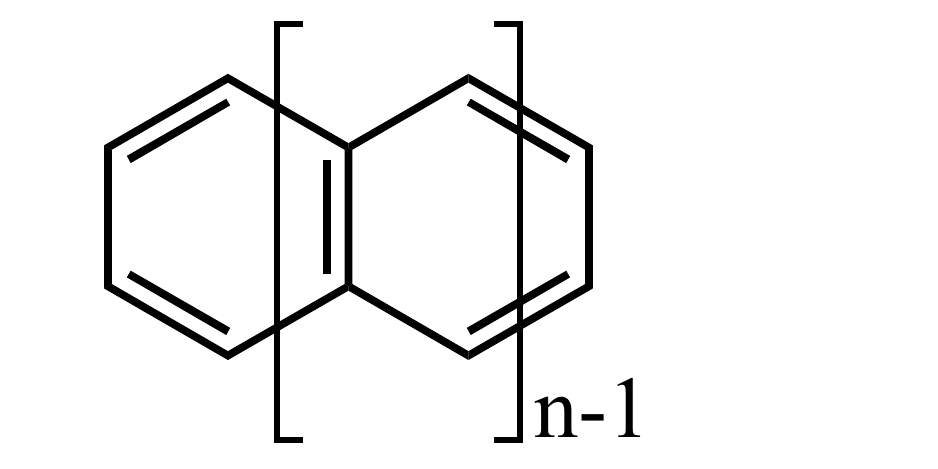}}
        \end{overpic}
        }
    \end{minipage}
    \caption{
        Relative energies on a series of challenging strongly-correlated systems.
        \textbf{a)} Energy deviations versus CCSD(T) for non-covalent interaction energies of 11 systems of the S22 dataset \cite{jureckaBenchmarkDatabaseAccurate2006, marshallBasisSetConvergence2011}.
        \textbf{b)} Detailed comparison of benzene dimer interaction energy across methods.
        \textbf{c)} Singlet-triplet energy gap in $n$-acene from naphthalene to hexacene.
        \textbf{d)} $n$-acene energy gap error to ZPVE corrected experimental results~\citep{anglikerElectronicSpectraHexacene1982c,birksPhotophysicsAromaticMolecules1970,burgosHeterofissionPentacenedopedTetracene1977a,schiedtPhotodetachmentPhotoelectronSpectroscopy1997,siebrandRadiationlessTransitionsPolyatomic1967}.
        Shaded region corresponds to typical experimental uncertainty: \qty{\pm1}{kcal/mol} for S22 (a) and acenes (c-d), and experimental uncertainty for benzene dimer (b).}
    \label{fig:relative_energies}
\end{figure*}

By combining these techniques, all crucial operations for VMC training are accelerated by $\O(\nel)$ as sketched in \cref{fig:overview}c and measured in \cref{fig:scaling}c.
Our empirical measurements show that our total runtime per step $\Ttot$ grows only $\Ttot \sim \nel^{2.3}$ up to 500 valence electrons, instead of $\Ttot \sim \nel^{3.0}$ for existing neural wavefunctions.
When comparing absolute runtimes for a 200 valence electron system, our approach yields 12$\times$ to 16$\times$ speedups over existing neural wave functions (\cref{fig:scaling}d) and even larger speedups for larger molecules.
Notably, these speedups are on top of the speedups obtained by the forward Laplacian~\citep{liComputationalFrameworkNeural2024}, a recent efficient method to evaluate the Laplacian of $\Psi$.
Thus, speedups are even greater compared to the original FermiNet~\citep{pfauInitioSolutionManyelectron2020,spencerBetterFasterFermionic2020} and Psiformer~\citep{vonglehnSelfAttentionAnsatzAbinitio2023} implementation.

\subsection{Accurate relative energies}
\label{sec:accurate_energies}
In the following, we demonstrate that FiRE not only accelerates NN-VMC but maintains high accuracy in various settings, such as non-covalent binding, singlet-triplet gaps, or ionization potentials.
We test these tasks on diverse systems, including biochemical compounds, hydrocarbons, and organometallic compounds.

\begin{figure*}[ht]
    \centering
    \begin{minipage}[t]{0.39\linewidth}
        \includegraphics[width=\columnwidth]{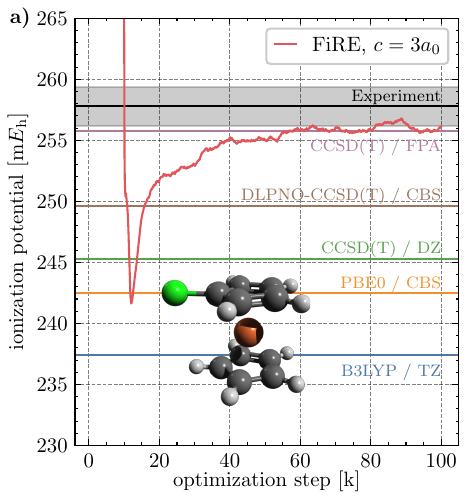}
    \end{minipage}
    \hfill
    \begin{minipage}[t]{.59\textwidth}
        \includegraphics[width=\columnwidth]{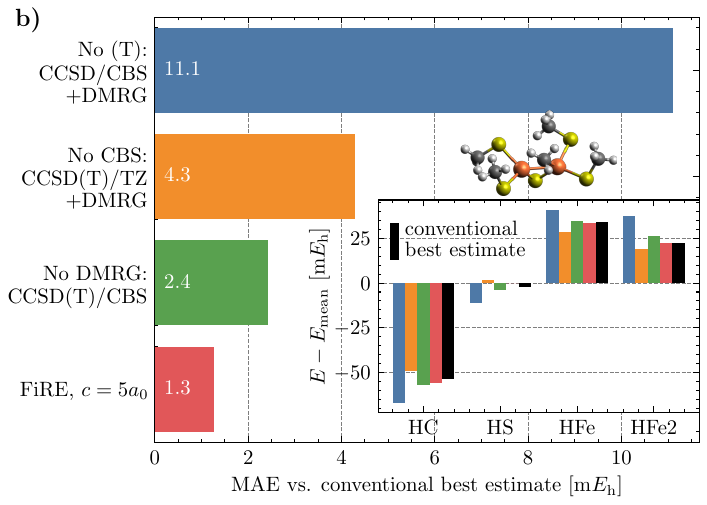}
    \end{minipage}
    \captionof{figure}{\textbf{Organometallic compounds}: \textbf{a)} Ionization potential of chloroferrocene as a function of optimization steps. \textbf{b)} Mean absolute error for protonation of iron-sulfur complex for conventional methods and FiRE. Inset: relative energies of 3 protonation sites vs HC site.}
    \label{fig:organometallic}
\end{figure*}

\paragraph{Non-covalent interactions} When restricting the range of electron embeddings, a natural question is how this affects the model's ability to capture long-range non-covalent interactions.
We investigate this behavior by comparing FiRE, LapNet~\citep{liComputationalFrameworkNeural2024}, and CCSD(T)/CBS interaction energies for 11 non-covalent interactions from the S22 dataset~\citep{jureckaBenchmarkDatabaseAccurate2006,marshallBasisSetConvergence2011}.
The systems include hydrogen bonds, dispersion energies, and mixed interactions.
For FiRE, we set the cutoff to $\cutoff=5a_0$ as determined by our ablation study in \cref{sec:cutoff}.
This is larger than the shortest distance between the interacting molecules but substantially smaller than the size of each entire complex.
Like \textcite{liComputationalFrameworkNeural2024}, we compare the energy of the bound system with the energy of the molecules separated by \SI{10}{\angstrom}.
The errors relative to CCSD(T) are plotted in \cref{fig:relative_energies}a.
It is apparent that FiRE accurately reconstructs the interaction energies, yielding a mean absolute error (MAE) of \qty{0.5}{\mHa} when using energy extrapolation (see \cref{sec:energy_extrapolation}) and \qty{2.3}{\mHa} without any extrapolation, compared to LapNet's \qty{2.3}{\mHa}.
Thus, even small cutoffs are sufficient for capturing complex long-range interactions.

Among the previous systems, the T-shaped benzene dimer is a particularly well-studied system where a variety of NN-VMC methods~\citep{renGroundStateMolecules2023,vonglehnSelfAttentionAnsatzAbinitio2023,liComputationalFrameworkNeural2024} attempted to reconstruct the experimental results by \textcite{groverDissociationEnergiesBenzene1987} and \textcite{krauseBindingEnergiesSmall1991}.
We compare all NN-VMC methods to the zero-point vibrational energy (ZPVE) corrected experimental results and CCSD(T)/CBS \cite{marshallBasisSetConvergence2011} in \cref{fig:relative_energies}b.
Furthermore, we study the cutoff effect by evaluating FiRE with $\cutoff\in\{3a_0,5a_0,7a_0\}$.
While previous works like \textcite{renGroundStateMolecules2023} overestimate the energy gap significantly, \textcite{vonglehnSelfAttentionAnsatzAbinitio2023}'s calculations are inconsistent in that the generally more accurate Psiformer significantly underestimates the gap compared to FermiNet.
For FiRE, results with all tested cutoffs are within the experimental uncertainty.
At $\cutoff=3a_0$ and $\cutoff=5a_0$ FiRE probably slightly underestimates the true interaction energy, yielding \qty{2.9}{\mHa} and \qty{3.6}{\mHa} respectively.
At $\cutoff=7a_0$ FiRE yields \qty{4.6}{\mHa}, which is in almost perfect agreement with both CCSD(T) (\qty{4.3}{\mHa}) and the ZPVE corrected experimental value of \qty{4.4}{\mHa} by \citeauthor{groverDissociationEnergiesBenzene1987}, which is considered to be the more accurate experiment~\citep{sinnokrotEstimatesInitioLimit2002,renGroundStateMolecules2023}.

\paragraph{Singlet-triplet gaps} Beyond interaction energies, we investigate the singlet-triplet gaps on a series of increasingly larger $n$-acenes from naphthalene (C$_{10}$H$_8$) to hexacene (C$_{26}$H$_{16}$).
Previous work found accurate methods such as CCSD(T)/FPA~\citep{hajgatoFocalPointAnalysis2011}, ACI-DSRG-MRPT2~\citep{schriberCombinedSelectedConfiguration2018}, and AFQMC~\citep{sheeSingletTripletEnergy2019} to be in disagreement with experimental results~\citep{anglikerElectronicSpectraHexacene1982c,birksPhotophysicsAromaticMolecules1970,burgosHeterofissionPentacenedopedTetracene1977a,schiedtPhotodetachmentPhotoelectronSpectroscopy1997,siebrandRadiationlessTransitionsPolyatomic1967}.
We demonstrate that larger cutoffs are unnecessary for covalently bound organic compounds, and $\cutoff=3a_0$ suffices to obtain accurate energies.
We obtain the respective states by setting the magnetic spin number $s_z=\Nup-\Ndown$ to 0 for singlet and 2 for triplet states and enforce state purity with the $S_+$ loss from \textcite{liSpinsymmetryenforcedSolutionManybody2024}.
For naphthalene and anthracene, energies converged well within 50k steps, and the remaining molecules were optimized for 100k steps.
The resulting state gaps of FiRE, AFQMC, CCSD(T), and ACI-DSRG-MRPT2, depending on the system size, are visualized in \cref{fig:relative_energies}c whereas \cref{fig:relative_energies}d shows the error relative to the ZPVE corrected experimental results.
Notably, CCSD(T) and AFQMC consistently overestimate the energy gap, whereas ACI-DSRG-MRPT2 underestimates it.
Despite the shrinking energy gaps in $n$, the reference methods' errors increase with system size. 
On the other hand, FiRE remains closest to the experimental results, exhibiting minimal deviations across all systems.
On average, FiRE's MAE to the experimental gap is \SI{1.7}{\milli\hartree}, whereas CCSD(T), ACI-DSRG-MRPT2, and AFQMC deviate by \SI{4.6}{\milli\hartree}, \SI{4.1}{\milli\hartree}, and \SI{4.4}{\milli\hartree}, respectively.
Interestingly, as seen in our ablations in \cref{sec:ablations}, our attention-based Jastrow factor, which introduces global-ranged correlation without affecting the overall scaling of FiRE (see \ref{sec:jastrow}), contributes substantially to this accuracy.

\paragraph{Organometallic compounds}
Organometallic compounds are of interest due to their widespread use in catalysis. However, their description is computationally challenging: methods, such as MRCI and CCSD(T), are either too costly to be applied in sufficiently large basis sets or require additional correction terms for an accurate assessment. 

As a first example, we compute the ionization potential (IP) of chloroferrocene, a known failure case of DFT~\citep{tomaIonizationEnergyReduction2020a}.
Like the singlet-triplet gaps, we set the FiRE cutoff to $\cutoff=3a_0$ because the system composes only short bond lengths.
The convergence of the IP in the number of optimization steps is visualized in \cref{fig:organometallic}a.
FiRE converges to an energy gap of \SI{256.5}{\milli\hartree} close to the experimental results of \SI{258}{\milli\hartree}~\citep{vondrakElectronicStructureHalogenoferrocenes1984}.
This agreement is unmatched by various other methods, e.g., DFT with B3LYP~\citep{inkpenUnusualRedoxProperties2015} deviates by \SI{20}{\milli\hartree}, and \SI{15}{\milli\hartree} with the PBE0 functional. CCSD(T) in a ccpvDZ basis -- the largest CCSD(T) calculation we could afford -- underestimates the IP by \SI{13}{\milli\hartree}, and the DLPNO-CCSD(T) approximation in the complete basis set limit (CBS) deviates by \SI{8}{\milli\hartree}.
Only when combining DLPNO-CCSD(T)/CBS energies with a CCSD(T) correction at the DZ level (denoted as CCSD(T)/FPA) do the energies match FiRE's accuracy.

Even more challenging is the protonation of the iron-sulfur complex \ch{[HFe_2S(CH_2)(SCH_3)_4]^{3-}}, which has been studied as a model system for catalysis in nitrogenase~\citep{zhaiMultireferenceProtonationEnergetics2023}. 
This iron-sulfur complex has four competing binding sites for an added proton: HC, HS, HFe, and HFe2.
\textcite{zhaiMultireferenceProtonationEnergetics2023} found that even CCSD(T) in the complete basis set limit (CBS) is insufficient to resolve the energy differences between these binding sites at chemical accuracy.
Their final best estimate is a compound estimate, requiring a relativistic coupled cluster calculation, perturbative triplets, CBS extrapolation, and estimation of multireference effects based on a separate DMRG calculation.
Omitting any of these corrections substantially increases the error, as depicted in \cref{fig:organometallic}b.
We use FiRE with $c=5a_0$ to account for larger bond lengths between the iron and sulfur cores.
Relativistic effects are part of the correlation consistent effective core potentials (ccECP)~\citep{bennettNewGenerationEffective2017} used throughout this work.
Unlike CCSD(T), FiRE does not require any corrections and still agrees with \citeauthor{zhaiMultireferenceProtonationEnergetics2023}'s compound estimate within chemical accuracy, with a mean absolute error of only \qty{1.3}{\mHa}, outperforming CCSD(T)/CBS which has a mean absolute error of \qty{2.4}{\mHa}.
With 180 electrons, this is not only the largest NN-VMC calculation done so far but also demonstrates the generality of FiRE even in cases where CCSD(T)/CBS does not achieve chemical accuracy.

Overall, we demonstrated that FiRE accurately describes non-covalent interactions, singlet-triplet gaps, and ionization potentials on various systems.
At this accuracy, it is unclear whether the remaining errors are due to errors in references, e.g., CCSD(T) errors, comparing \SI{0}{\kelvin} gas phase to experimental conditions, or structural relaxations which may affect relative energies~\citep{schriberCombinedSelectedConfiguration2018}.

\subsection{Convergence rates for NN-VMC}
\label{sec:convergence_rates}
Our ability to optimize neural wave functions for such large systems enables us to study the scaling behavior of NN-VMC for the first time.
When analyzing the errors in absolute energies for acenes (\cref{sec:accurate_energies}) and  cumulenes (\cref{sec:cumulene}), as a function of system size $\nel$ and number of optimization steps $t$, we find good agreement with a power law of the form
\begin{equation}
    E(t, \nel) - E(\infty, \nel) \propto t^{-\alpha} \nel^\beta,
\end{equation}
as depicted in \Cref{fig:scaling_laws}.
Interestingly, we find similar exponents of $\alpha\approx 1$ and $\beta \approx 2.3$ across systems.
Some recent theoretical work on convergence rates of VMC has also obtained polynomial convergence in the number of steps, although at lower rates \cite{abrahamsenConvergenceVariationalMonte2024,liConvergenceAnalysisStochastic2024}. 
While their analysis is not directly applicable to our setting, we give a short comparison in \cref{sec:theoretical_convergence_rates}.
Extrapolating from our empirical rates, to reach a given error in absolute energy, the number of optimization steps needs to scale as $t \sim \nel^\frac{\beta}{\alpha} \approx \nel^{2.3}$.

\begin{Figure}
\captionsetup{type=figure}
\includegraphics[width=\columnwidth]{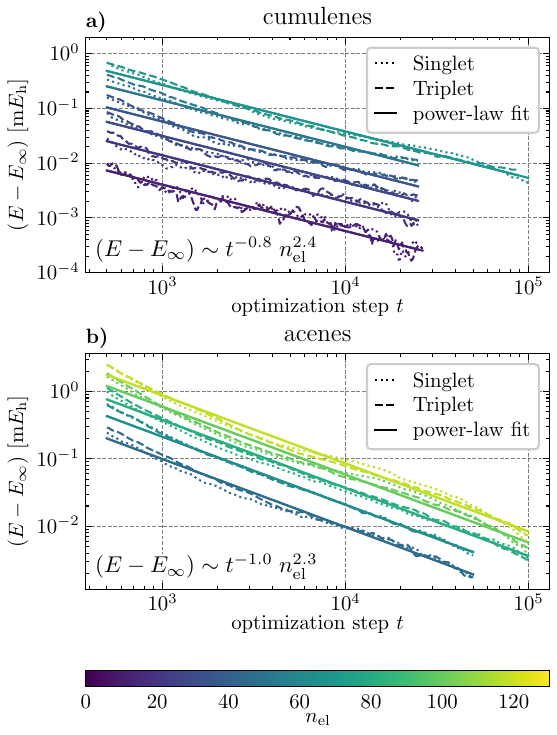}
\captionof{figure}{\textbf{Convergence rates for neural wave functions}: Absolute energy error as a function of optimization steps for molecules of increasing size: \textbf{a)} cumulenes \textbf{b)} acenes. For both systems, the optimization curves are well approximated by a powerlaw with similar exponents.}
\label{fig:scaling_laws}
\end{Figure}

\section{Discussion}
We have pushed the boundary of NN-VMC in the two most important dimensions: efficiency and accuracy.
Unlike traditional VMC, FiRE qualitatively and quantitatively reconstructs experimental results on various challenging systems, even in cases where contemporary NN-VMC disagrees.
At the same time, FiRE accelerates NN-VMC by $\O(\nel)$, yielding speedups of up to an order of magnitude for the systems investigated.
With FiRE, we obtain highly accurate energies for system sizes, which become inaccessible to many high-accuracy methods, and the remaining ones require expert knowledge to be applied correctly.
Compared to methods like MR-CI or CCSD(T), our NN-VMC works out of the box and requires no method combinations, basis sets, or active space, reducing the need for expert knowledge for high-accuracy quantum chemistry.
Furthermore, unlike other methods, NN-VMC yields accurate energies and provides the corresponding wave function, thus giving, in principle, access to any ground-state property.
While NN-VMC has so far rarely been applied to practical chemical problems, we firmly believe FiRE is fast and accurate enough to earn a place in the practitioner's toolbox.

Still, open questions and challenges remain.
While we obtain state-of-the-art results for several systems containing a variety of non-local interactions, the assumptions of our FiRE may fail for some classes of systems.
Compared to dense NN-VMC, FiRE's absolute energies are less accurate when choosing an aggressive cutoff.
In agreement with previous works~\citep{vonglehnSelfAttentionAnsatzAbinitio2023}, we observed that larger systems require more optimization steps, an issue that is not unique to NN-VMC -- conventional methods are also increasingly complicated to converge with increasing system size and require careful tuning of optimization parameters~\citep{lehtolaOverviewSelfConsistentField2020}.
FiRE also adds some implementation complexity compared to dense NN-VMC because implementations for low-rank updates, sparse forward-mode Laplacian computations, and padding for GPUs are necessary (\cref{sec:jax_implementation}).
Finally, while in the limit of many electrons, the scaling remains the same for periodic or bulk systems; densely packed structures increase the neighborhood size, yielding higher compute times.

We expect future work to investigate these aspects and further improve our approach's accuracy and compute time, at last, by transferring deep learning advancements to ab-initio quantum chemistry.
Further, we hope that our quantitative convergence rate results serve as the basis for further research into how NN-VMC scaling depends on system properties, such as the spectral gap, and optimization choices, such as preconditioning and learning rate scheduling.
Beyond the study of gas-phase molecules, we expect FiRE to accelerate progress in various fields of the physical sciences as it is directly applicable to the many domains in which NN-VMC has shown early promise, such as photochemistry~\citep{pfauAccurateComputationQuantum2024}, solid-state physics~\citep{liInitioCalculationReal2022,gerardTransferableNeuralWavefunctions2024}, nuclear physics~\citep{lovatoHiddennucleonsNeuralnetworkQuantum2022}, positron chemistry~\citep{cassellaNeuralNetworkVariational2023}, polaritonic chemistry~\citep{tangDeepQuantumMonte2025} or the study of topological materials \cite{liDeepLearningSheds2025}.

\section{Methods}
\subsection{Variational Monte Carlo}
We seek to solve the stationary Schrödinger equation within the Born-Oppenheimer approximation
\begin{align}
  \H\ket{\wf} = E\ket{\wf} \label{eq:schrodinger}
\end{align}
where $\H: \mathcal{H}^2\to\mathcal{L}^2$ is the Hamiltonian operator and $\wf:\R^{\nel \times 3}\to\R$ is the electronic wave function. 
Here, we follow standard practice and use a spin-assigned wavefunction where the first $\Nup$ electrons are spin-up and the latter $\nel-\Nup$ are spin-down.
In atomic units, the Hamiltonian for a molecular system is given by
\begin{align}
\begin{split}
  \H :=& -\frac{1}{2}
    \sum_{i=1}^{\nel}\sum_{k=1}^{3}\frac{\partial^2}{\partial \r_{ik}^2}
    + \sum_{j>i}^{\nel} \frac{1}{| \epos_i - \epos_j |}\\
    &- \sum_{i=1}^{\nel}\sum_{m=1}^{\Nnuc} \frac{\charge_m}{| \epos_i - \npos_m |}
    + \sum_{n>m}^{\Nnuc} \frac{\charge_m \charge_n}{| \npos_m - \npos_n |}.
\end{split}\label{eq:hamiltonian}
\end{align}
We assume that the minimum of the spectrum of $\H$ is given as an isolated eigenvalue $E_0$ of finite multiplicity, which we call the \emph{ground-state energy} and corresponding eigenfunctions are referred to as \emph{ground states}. 
To compute ground states/ground-state energies and solve Equation~\ref{eq:schrodinger}, we seek to minimize the variational energy
\begin{align}
    E\left[\Psi\right] &= \frac{\bra{\wf}\H\ket{\wf}}{\langle\wf\vert\wf\rangle} = \Ewf\underbrace{\left[
        \wf(\rvec)^{-1} [\H \wf](\rvec)
        \right]
  }_{\El(\rvec)}\geq E_0,\label{eq:expectation}
\end{align}
where $\wfp(\rvec)=\frac{\wf(\rvec)^2}{\langle\wf\vert\wf\rangle}$. By the Raleigh-Ritz principle, upper bounds the ground-state energy $E_0$.
To compute $E[\wf]$, we use importance sampling to evaluate the expectation in \cref{eq:expectation} using the Metroplis-Hastings algorithm.
The so-called local energy $\El(\rvec)$ can be computed via
\begin{align}
  \begin{split}
    \El(\rvec) =& \underbrace{
      -\frac{1}{2} \left(\Delta\lwf+\left(\nabla\lwf\right)^2\right)
    }_\text{kinetic energy} + V(\rvec)
  \end{split}
\end{align}
where $V$ is the potential energy, i.e., the last three terms in \cref{eq:hamiltonian}.
Note that we, in practice, use pseudopotentials as described by \textcite{liFermionicNeuralNetwork2022}.

We aim to approximate $E_0$ by minimizing $\theta \mapsto E[\wf_\theta]$ over a parametrized class $\{\wf_\theta\}$ of (neural network-based) wave functions. To this end we use gradient-based optimization
\begin{align}
    \theta^{t+1} = \theta^t - \eta^t \delta^{t}
\end{align}
with learning rate $\eta^t\in\R_+$ and update $\delta^t$.
While one could naively use the gradient of the energy
\begin{align}
    \nabla_\theta E[\wf] \propto \Ewf\left[
      \left(\El(\rvec) - \Ewf\left[\El(\rvec)\right]\right)\glwf
    \right]
\end{align}
as the update, quasi-Newton optimizers yield faster convergence.
Thus, we use the stochastic reconfiguration-inspired SPRING algorithm \cite{rendeSimpleLinearAlgebra2023, goldshlagerKaczmarzinspiredApproachAccelerate2024a} to obtain the parameter updates
\begin{align}
  \delta^t = \mcqgt \left(\mcqgt^T\mcqgt + \lambda I\right)^{-1}\left(\epsilon - \mcqgt\eta\delta^{t-1}\right) + \eta\delta^{t-1}\label{eq:spring}
\end{align}
where $\mcqgt_i=\nabla_\theta\ln\wf(\rvec^{(i)}) - \frac{1}{N}\sum_{j=1}^N\nabla_\theta\ln\wf(\rvec^{(j)})$ and $\epsilon_i=\El(\rvec^{(i)})-\frac{1}{N}\sum_{j=1}^N\El(\rvec^{(j)})$ for a batch of $N$ samples $\rvec^{(i)}\sim\wfp$. This essentially corresponds to a numerical approximation of stochastic reconfiguration/natural gradient descent $\delta_t = \Ewf\left[\nabla_\theta\ln\wfp(\rvec)\nabla_\theta\ln\wfp(\rvec)^T\right]^{-1}\nabla_\theta E[\wf]$ with momentum.

\subsection{Wave function ansatz}\label{sec:model}
As alluded to in the introduction, our wave function follows the form of neural-network Slater-Jastrow wave functions like \cref{eq:slater_jastrow} with a linear combination of a small number of determinants
\begin{equation}
    \Psi(\rvec) = \Jastrow(\rvec) \sum_{d=1}^{\ndet} \det [\Phi_d(\rvec)]. \label{eq:wf}
\end{equation}
The entries of the orbital matrices $\Phi$ do not depend on just a single electron $\rvec_i$, but instead on a so-called embedding vector $\vh_i$ (see \cref{eq:orbitals}), which represents the electron $i$ and its environment.
The Jastrow factor $\Jastrow$ further includes range-unlimited electron correlation effects.

\paragraph{Finite-range embeddings}
The efficiency of our neural wave function ansatz rests on the locality assumption of electron correlation effects.
As mentioned above, we construct the wave function's electron embeddings $\vh_i$ such that it only depends on $\{\epos_j: |\epos_j - \epos_i|\leq c\}$ for some cutoff $c$.
We accomplish this with a modified version of \textcite{gaoGeneralizingNeuralWave2023}'s graph neural network-like ansatz.
Before detailing the architecture, we define pairwise features $\ve_{ij} \in \R^4$ for pairs of electrons and $\hat{\ve}_{im} \in \R^4$ for electron-nucleus pairs:
\begin{align}
    \ve_{ij} &= \text{Concat}[\vert\epos_i-\epos_j\vert,\epos_i-\epos_j],\\
    \hat{\ve}_{im} &= \text{Concat}[\vert\epos_i-\npos_m\vert,\epos_i-\npos_m].
\end{align}
We start with constructing initial electron embeddings $\vh^0_i$ given the nuclear position $\nposs$ and charges $\charges$, i.e., independent of all other electrons:
\begin{align}
    \vh_i^0 =& \text{GLU}\left(
        \sum_{m=1}^\Nnuc \Gamma_m(\hat{\ve}_{im}) \odot \left(
            \nucemb_m + \hat{\bar{\ve}}_{im}W
        \right)
    \right).
\end{align}
Here, $\hat{\bar{\ve}}_{im}$ is a rescaled electron-nuclei distance vector 
\begin{equation}
    \hat{\bar{\ve}}_{im}=\frac{\log(1+|\rvec_i-\npos_m|)}{|\rvec_i-\npos_m|}\hat{\ve}_{im},
\end{equation}as proposed by \cite{vonglehnSelfAttentionAnsatzAbinitio2023}, and $\text{GLU}$ is a gated linear unit~\citep{shazeerGLUVariantsImprove2020} with LayerNorm~\citep{baLayerNormalization2016} as common in contemporary deep learning~\citep{touvronLLaMAOpenEfficient2023}. 
The vector $\nucemb_m \in \R^d$ is a trainable embedding representing the $m$th nucleus and $\Gamma_m:\R^4\to\R^d$ is a spatial filter of the $m$th nucleus that featurizes the distance and ensures a smooth decay to 0 at $\ncutoff$, i.e., $x_0\geq \ncutoff\implies \Gamma_m(\vx)=\mathbf{0}$:
\begin{align}
    \Gamma_m(\vx) &= \sigma(\vx \mW_m + \vb_m)\mW \odot \chi(x_0)\mW_\text{env},\\
    \chi(x)&= f_\text{cut}(x)\odot \text{Concat}[\exp(-\sigma_i^2x^2)]_{i=1}^{d_0}
\end{align}
where $\chi:\R_+\to\R^{d_0}$ is a set of nuclei-centered Gaussian multiplied with the polynomial cutoff function $f_\text{cut}:\R_+\to\R_+$ function from \textcite{gasteigerDirectionalMessagePassing2021}.
The parameters $\sigma_i$ control the width of additional Gaussian envelope functions.
This way, the wave function is smooth if an electron moves in or out of the cutoff range. 
Next, we update the $i$th electron based on the embeddings of all electrons within the cutoff $c$ by performing a single message passing step to obtain
\begin{align}
    \vh^1_i =& \vh^0_i + \vm^\parallel_i + \vm^\nparallel_i, \\
    \vm^\alpha_i =& \sum_{j\in\N^\alpha_{\epos_i}}\Gamma(\ve_{ij})\odot\sigma\left(\text{Concat}[\vh^0_i,\vh_j^0,\bar{\ve}_{ij}]\mW + \vb\right)
\end{align}
with $\Gamma$ of the same form as the $\Gamma_m$ above but without a dependence on any nucleus and $\cutoff$ instead of $\ncutoff$. 
Here, $\N^\alpha_{\epos_i}$ denotes the set of electron indices that are within cutoff $\cutoff$ and have either parallel $\alpha=\parallel$ or opposing spin $\alpha=\nparallel$. The features $\bar{\ve}_{ij}$ are rescaled electron-electron distance vectors analogous to $\hat{\bar{\ve}}_{im}$.
Finally, we apply a multi layer perceptron (MLP) to the electron embeddings
\begin{align}
    \vh_i = \text{MLP}(\vh^1_i).
\end{align}
We purposefully avoid multiple rounds of message passing as this would introduce costly long-range dependencies at diminishing returns~\cite{gaoGeneralizingNeuralWave2023}.
Instead, we recommend increasing the cutoff when higher accuracy is required.

From these electron embeddings, we construct orbitals via linear projections and envelopes $\varphi_l:\R^3\to\R$ which ensure exponential decay (as known to hold for ground states \cite{agmon2006bounds}):
\begin{align}
    \Phi_{dil} = \vh_i \mW_{dl} \odot \varphi_l(\epos_i).
\end{align}
For the envelopes, we use the improved exponential envelopes from \textcite{gaoNeuralPfaffiansSolving2024}:
\begin{align}
    \varphi_l(r)=\sum_{m=1}^\Nnuc \sum_{e=1}^\Nenv \pi_{lme} e^{\sigma_{me}| r - \npos_m|}.
\end{align}

\paragraph{Global electron correlation effects}\label{sec:jastrow}
Beyond the finite-range multi-electron orbitals, which capture correlation effects within the cutoff range, our ansatz contains several mechanisms to capture global electron correlations.
Our ansatz is the sum of a small number of determinants (typically $N_\text{det}=4$), which captures static correlation, see \cref{eq:wf}.
To capture dynamic correlation, we additionally use a 3-term permutation-symmetric Jastrow factor:
\newcommand{\jcusp}{\Jastrow_\text{cusp}}
\newcommand{\jmlp}{\Jastrow_\MLP}
\newcommand{\jatt}{\Jastrow_\text{att}}
\begin{align}
\Jastrow(\rvec) &= \jcusp(\rvec) + \jmlp(\rvec) + \jatt(\rvec).
\end{align}
To enforce the electronic cusp conditions, we use \textcite{vonglehnSelfAttentionAnsatzAbinitio2023}'s cusp Jastrow factor
\begin{align}
    \begin{split}
    \jcusp(\rvec) =& \exp\left(\sum_{\substack{0<i<j\leq\Nup,\\ \Nup < i<j\leq \nel}} \frac{\omega_\text{par}\alpha_\text{par}^2}{\alpha_\text{par}+ |\epos_i-\epos_j|}\right.\\
    &+ \left.\sum_{0<i\leq\Nup<j\leq \nel} \frac{\omega_\text{anti}\alpha_\text{anti}^2}{\alpha_\text{anti} + |\epos_i-\epos_j|}\right)
    \end{split}
\end{align}
with learnable parameters $w_\text{par},w_\text{anti},\alpha_\text{par},\alpha_\text{anti}\in\R$.
In addition to this constrained Jastrow factor, we add two neural network-based Jastrow factors.
The first one is a per-electron MLP-based Jastrow factor
\begin{align}
    \jmlp(\rvec) &= \exp\left(\sum_{i=1}^\nel\MLP_1(\vh_i)\right) \left(\sum_{i=1}^\nel \MLP_2(\vh_i)\right)
\end{align}
from \textcite{gaoSamplingfreeInferenceInitio2023} where in addition to the log readout via $\MLP_1$, we add a node-inducing component via $\MLP_2$.
Note that the MLPs in this Jastrow factor still only have access to the local environment of individual elements, with the total Jastrow factor being the sum of the individual electrons, limiting the correlation effects that can be captured.

To capture global electron correlation effects, one could apply an MLP to an average electron embedding, but such a Jastrow factor would lose access to high-frequency information due to the averaging.
Instead we propose a novel Jastrow factor based on cross attention to so-called registers~\citep{darcetVisionTransformersNeed2023}.
For each register $r\in\{1,\ldots,\Nreg\}$, we define a query $\vq_r\in\R^{D}$, and weights $\mW^V_r\in\R^{D\times \dreg}$. We perform cross attention between the electron embeddings $\mH \in \R^{\nel \times D}$ (used as keys) and the register queries to obtain the register embeddings
\begin{align}
    \vv_r = \softmax\left(\mH \vk_r\right)^T\mH\mW^V_r.
\end{align}
Similar to the per-electron MLP Jastrow factor, we perform a 2-step readout on $\mV=\text{Concat}[\vv_r]_{r=1}^{\Nreg}\in\R^{\Nreg\dreg}$:
\begin{align}
    \jatt(\rvec) = \exp\left(\MLP_1\left(\mV\right)\right) \MLP_2\left(\mV\right).
\end{align}
We demonstrate the importance of this Jastrow factor in \cref{sec:ablations}, where we observe an approximately \qty{10}{\mHa} improvement of absolute energies and $\qty{2}{\mHa}$ for relative energies.
Note that while this Jastrow factor can capture correlation between electron embeddings at arbitrary distance, a forward pass through this Jastrow factor scales linearly with the number of electrons and therefore does not affect overall scaling of our method.

\subsection{Low-rank wave function updates}\label{sec:method_low_rank_update}
Updates of the wavefunction after moving a small number of electrons are the key step for Monte Carlo sampling or evaluation of pseudopotentials.
To enable these efficient low-rank updates, we store all intermediate embeddings of the network when computing $\Psi$.
When changing the positions of $K<\nel$ electrons with indices $\iota_1,\ldots,\iota_K$ to positions $\hat{\rvec}_{\iota_1},\ldots,\hat{\rvec}_{\iota_K}$, we determine the update set $\updset=\{\iota_1,\ldots,\iota_K\}\cup\bigcup_{k=1}^K\N_{\rvec_{\iota_k}} \cup \N_{\rvec'_{\iota_k}}$ of all electron indices $i$ which are within the cutoff $\cutoff$ of a moved electron's previous or new location.
For any physically plausible molecule Coulomb repulsion spreads the electrons across the molecule, such that the average number of electrons within a given cutoff radius does not scale with system size.
Therefore, given a large enough system, the size of the update set $|\updset|$ is independent of $\nel$.
We then only recompute the embeddings $\vec{h}_i$ for this bounded number of electrons $i \in \updset$. 
The same technique is applied to all other parts of the wave function, such as the Jastrow factor.

Of particular importance is the update of the determinant in \eqref{eq:slater_jastrow}.
If only the rows $i \in \updset$ of $\Phi$ are changed, we can express the resulting orbital matrix $\Phi'$ as a low rank update
\begin{equation}
    \Phi' = \Phi + \sum_{i \in \updset}\vec{e}_i (\Phi'_i - \Phi_i)^T = \Phi + \mU \mV^T.
\end{equation}
Here, $\vec{e}_i \in \mathbb{R}^\nel$ denotes the $i$th unit vector, and $\mU, \mV \in \mathbb{R}^{\nel \times |\updset|}$ denote the matrices of all unit vectors $\vec{e}_i$ and changes in the orbital matrix $V_i = (\Phi'_i - \Phi_i)$ for $i \in \updset$.
Like conventional VMC~\citep{mcdanielDelayedSlaterDeterminant2017}, the matrix determinant lemma and the Woodbury matrix identity enable us to compute updates for $\det [\Phi]$ and $\Phi^{-1}$
\begin{align}
    \mA :=& (I_{|\updset|} + \mV^T \Phi^{-1} \mU), \\
    \det [\Phi'] =& \det \left[\Phi + \mU \mV^T\right] = \det[\Phi] \det[\mA],  \\
    \Phi'^{-1} =& \Phi^{-1} - \Phi^{-1} \mU \mA^{-1} \mV \Phi^{-1}.
\end{align} Constructing $\mA$ requires $\O(\nel\cdot|\updset|^2)$ operations and since $\mA$ is only in $\mathbb{R}^{|\updset|\times|\updset|}$, computing its inverse and determinant only scales as $\O(|\updset|^3)$.
We note that we can compute the determinant and the inverse of $\mA$ from the same LU decomposition, requiring low additional computational effort.
The same idea can also be used when computing the original full inverse $\Phi^{-1}$ from the LU obtained during determinant computation.

We use low-rank updates during Monte Carlo sampling and when evaluating non-local operators such as the effective core potential and the $S^+$ spin operator.
All of these require evaluating ratios of the form $\wf(\rvec')/\wf(\rvec)$.
For single electron moves during Monte Carlo sampling and the non-local effective core potential, $\rvec'$ and $\rvec$ differ in only a single electron.
For the $S^+$ operator, they differ in only two electrons.

\subsection{Efficient Laplacian}\label{sec:Laplacian}
Our finite-range embeddings allow for a more efficient computation of the Laplacian of the wave function, which is required for the kinetic energy.
For a composite function $f=f_N\circ\ldots\circ\ f_1$ of some input $\vx_1\in\R^{d_1}$, the forward Laplacian framework \cite{liComputationalFrameworkNeural2024} propagates the primal $\vx_i\in\R^{d_i}$, the Jacobian $\nabla \vx_i\in\R^{d_i\times d_1}$, and the Laplacian $\Delta \vx_i\in\R^{d_i} $:
\begin{align}
    \vx_{i+1} &= f_i(\vx_{i}),\\
    \nabla \vx_{i+1} &= J^{f_i}(\vx_i) \nabla\vx_i, \label{eq:forward_lap_jac}\\
    \Delta \vx_{i+1} &= J^{f_i}(\vx_i) \Delta\vx_i + \Tr\left[(\nabla\vx_i)^T H^{f_i}(\vx_i)  \nabla \vx_i\right] \label{eq:forward_lap_lap}
\end{align}
where $J^{f_i}(\vx_i)$ and $H^{f_i}(\vx_i)$ are the Jacobian and Hessian of $f_i$ at $\vx_i$.
Most of the computation is here frequently dominated by the propagation of the Jacobian $\nabla\vx$ which scales linearly with the domain of $f$ and the computation of $\Tr\left[(\nabla\vx_i)^T H^{f_i}(\vx_i)  \nabla \vx_i\right]$.

Our local updates accelerate the Laplacian computation due to sparse Jacobians $\nabla\vh_i$ as an electron's embedding only depends on the electrons in its vicinity.
This way, we avoid materializing the full Jacobian but instead propagate sparse tensors, reducing the Jacobian propagation costs by $\O(\nel)$.
The case of the determinant is particularly noteworthy. 
The Jacobian and Hessian of the logarithm of the determinant are given as
\begin{align}
    J^{\ln\det}_{ij}(\Phi) &= \Phi^{-1}_{ji} \\
    H^{\ln\det}_{ij,km}(\Phi) &= - \Phi^{-1}_{jk} \Phi^{-1}_{mi}.
\end{align}
To compute the forward propagations in \cref{eq:forward_lap_jac,eq:forward_lap_lap}, we define the tensor $\vec{M} \in \mathbb{R}^{\nel \times \nel \times \nel \times \ndim}$ as the product of the Jacobian of the orbital matrix with its inverse
\begin{equation}
    M_{ik,nd} = \sum_j (\nabla_{nd} \Phi_{ij}) \Phi^{-1}_{jk}. \label{eq:forward_lap_det_M}
\end{equation}
The required terms for \cref{eq:forward_lap_jac,eq:forward_lap_lap} are then given as
\begin{align}
    J^{\ln\det}(\Phi) \nabla\Phi &= \sum_i M_{ii,nd} \\
    J^{\ln\det}(\Phi) \Delta\Phi &= \sum_{ij} \Delta\Phi_{ij} \Phi^{-1}_{ji} \\
    \Tr\left[(\nabla\Phi)^T H^{\ln\det}(\Phi)  \nabla \Phi\right] &= -\sum_{d=1}^\ndim \sum_{i,k,n=1}^{\nel} M_{ik,nd}M_{ki,nd}.
\end{align}
For fully correlated orbitals, the last sum contains $\nel^3$ terms for each combination of the indices $i,k,n$.
For finite-range orbitals, however, we can utilize the fact that $M_{ik, and}=0$ if $n \notin \nbset_i$, because in that case, the Jacobian for electron $i$ w.r.t. electron $n$ is zero.
Therefore, we can restrict this sum to
\begin{equation}
    \sum_{i,k,n=1}^{\nel} M_{ik,nd}M_{ki,nd} = \sum_{n=1}^\nel \sum_{i,k \in \nbset_n} M_{ik,nd}M_{ki,nd},\label{eq:logdet_lowrank}
\end{equation}
which reduces the complexity of this contraction from $\O(\nel^3)$ to $\O(\nel\nnb^2)$.
Another large advantage of range-limited orbitals arises in \cref{eq:forward_lap_det_M}. 
For fully correlated orbitals, this contraction has complexity $\O(\nel^4)$ since each of the $\O(\nel^3)$ entries of $M$ is a contraction over dimension $\nel$.
However, for finite-range embeddings, the Jacobian $\nabla_{nd} \Phi_{ij}$ is sparse, thus yielding corresponding sparsity in $M$, reducing the memory and compute cost by $\O(\nel)$.

\subsection{Improved Monte Carlo sampling}
\label{sec:methods_mcmc}
We use the Metropolis-Hastings algorithm \cite{metropolisEquationStateCalculations1953} to sample electron coordinates $\rvec$ from the wavefunction $\wf$.
The standard proposal distributions $\rho(\eposs'\vert\eposs)$ in NN-VMC propose new electron positions by perturbing the previous electronic coordinates with noise $\rho(\eposs'\vert\eposs)=\mathcal{N}(\eposs'\vert\eposs,\sigma^2 I)$.
While working well in covalent systems, it may lead to non-variational energies in largely separated sub-systems.
If the gap between two sub-systems is too large, the probability of moving an electron from one sub-system to the other decays to zero due to the exponential envelopes.
In such cases, the samples may not represent the wave function's distribution well.

We propose to additionally use global single-electron jumps to eliminate this issue.
While non-local moves have a history in diffusion Monte Carlo (DMC)~\citep{casulaLocalityApproximationStandard2006,casulaSizeconsistentVariationalApproaches2010}, there, they obey a specific form that ensures correct convergence but is costly to evaluate.
In contrast, VMC's proposal distribution's support set must only cover the target distribution's support set.
Thus, we define a Gaussian Mixture Model (GMM) proposal distribution
\begin{align}
    \rho_\text{global}(\epos')=\frac{1}{\sum_{m=1}^\Nnuc \charge_m}\sum_{m=1}^\Nnuc \charge_m\, \mathcal{N}(\epos'\vert\npos_m,\globalstep^2 I).
\end{align}
Unlike local Gaussian moves, these global moves do not have a symmetrical proposal distribution, and we, therefore, need to adjust the acceptance probability by a factor of 
\begin{equation}
    \frac{\rho(\rvec|\rvec')}{\rho(\rvec'|\rvec)} = \frac{\rho(\rvec)}{\rho(\rvec')},
\end{equation}
to obtain unbiased estimates.
While one traditionally optimizes the proposal distribution to yield an acceptance ratio of $\approx 50\%$ by adjusting the scale parameter $\sigma^2$ on the fly, i.e., with lower $\sigma^2$ yielding higher acceptance ratios as $\sigma^2\to0\implies\frac{\wf^2(\eposs')}{\wf^2(\eposs)}\to 1$, the same cannot trivially be done with $\rho_\text{global}$.
Due to the magnitude of the perturbation, we observe lower acceptance ratios for these global moves.
To set $\globalstep$, we compared acceptance ratios for $\globalstep\in\{1,2,3\}$ on the benzene dimer and chose $\globalstep=2$, which yielded the highest acceptance ratio of $\approx10\%$.
We find that this acceptance rate depends weakly on system size, ranging from 12\% for \ch{C_4H_4} to 6\% for \ch{C_{16}H_4}.
To maximize computational efficiency, we alternate between traditional single-electron moves, perturbing a single electron's position with noise, and global single-electron jumps.

\section{Code availability}
All code and data will be made openly available upon publication.
%
%

\section*{Acknowledgements}
We greatly appreciate Gunnar Arctaedius' and Leon Gerard's support on prototyping this approach.
This work has been funded by the Austrian Science Fund FWF Project I 3403, the WWTF Project ICT19-041 and the Federal Ministry of Education and Research (BMBF) and the Free State of Bavaria under the Excellence Strategy of the Federal Government and the Länder.
Computations were achieved with the Vienna Scientific Cluster, Leonardo (Project L-AUT 005) and the Munich Center for Machine Learning Cluster (MCML) hosted at the Leibniz Supercomputing Centre (LRZ).
The funders had no role in study design, data collection and analysis, decision to publish or preparation of the manuscript.
\section*{Author contributions statement}
MS and NG jointly conceived the project, implemented the approach, ran the experiments, and analyzed the data.
MS focused on the sparse forward Laplacian, MCMC, efficient JAX implementation, finding test systems, and running reference calculations.
NG focused on the architecture, low-rank updates, pseudopotentials, $S_+$ operator, and optimization.
MS, NG and PG wrote the paper.
SG and PG provided funding and feedback on the manuscript.
%

\printbibliography

\end{multicols}
\clearpage

\appendix
\renewcommand{\thefigure}{S\arabic{figure}}
\renewcommand{\thetable}{S\arabic{table}}
\renewcommand{\theequation}{S\arabic{equation}}
\renewcommand{\thesubsection}{\Alph{subsection}}
\setcounter{figure}{0}
\setcounter{table}{0}
\setcounter{equation}{0}
\newrefsection

\titleformat{\section}[display]
  {\normalfont\LARGE\bfseries\centering}  
  {}                                      
  {0pt}                                  
  {}                                     
  []                                     

\section{Supplementary Information}
\begin{center}
    \Large Accurate Ab-initio Neural-network Solutions to\\Large-Scale Electronic Structure Problems
\end{center}

\subsection{Effect of cutoff: \texorpdfstring{\ch{H_10}}{H10}}\label{sec:cutoff}
\begin{figure*}[!ht]
    \centering
    \includegraphics[width=\textwidth]{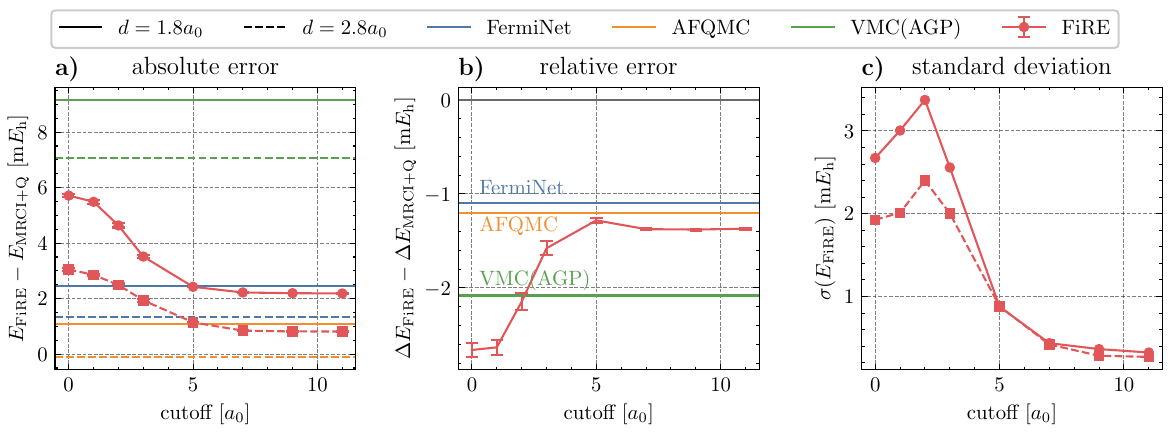}
    \caption{\textbf{Impact of cutoff on accuracy for \ch{H_{10}} hydrogen-chain} \textbf{a)} error in absolute energy relative to MRCI+Q \textbf{b)} error in relative energy $\Delta E=E_{2.8}-E_{1.8}$ relative to MRCI+Q \textbf{c)} standard deviation of the sampled energies}
    \label{fig:hchain_cutoff}
\end{figure*}

To investigate the effect of the cutoff radius $\cutoff$ on accuracy, we compute energies for \ch{H_{10}} hydrogen chains at inter-atomic spacings $d=\qty{1.8}{\bohr}$ and $d=\qty{2.8}{\bohr}$.
This toy system has been used to benchmark many high-accuracy methods \cite{mottaSolutionManyElectronProblem2017} because, despite its small size, even methods like CCSD(T) miss relative energies by up to \qty{15}{mHa}.
\Cref{fig:hchain_cutoff} depicts the errors in absolute energy, relative energy, and energy standard deviation.
We find that all quantities rapidly converge with increasing cutoff, reaching convergence at $\cutoff\approx\qtyrange{3}{5}{\bohr}$, which is much smaller than the length of the molecule (\qtyrange{16}{26}{\bohr}).
We also find our energies to be in good agreement with other high-accuracy methods, like FermiNet \cite{pfauInitioSolutionManyelectron2020} and AFQMC \cite{mottaSolutionManyElectronProblem2017}.
For cutoffs $\cutoff\geq\qty{5}{\bohr}$, we even obtain lower absolute energies than FermiNet despite being range-limited and having fewer determinants.
For the impact of hyperparameters other than the cutoff, see the ablation study in \cref{sec:ablations}.

Notably, in a densely packed system and for a sufficiently large $\nel$, the average number of neighbors of any electron $\nnb$ scale linearly in the volume, i.e., $\O(\cutoff^3)$.
Consequently, the wave function update sacles $\O(\cutoff^9)$.
We fix the cutoff to $\cutoff=3a_0$ for comparing ionization potentials and singlet-triplet gaps and $\cutoff=5a_0$ when computing interaction energies to optimize the tradeoff between compute time and accuracy.
We found this to be a favorable tradeoff between the accuracy of relative energies and compute time.

\subsection{Non-local interactions in hydrocarbons: cumulene}
\label{sec:cumulene}

\begin{figure*}[ht]
    \centering
    \begin{minipage}[t]{0.7\textwidth}
        \centering
        \includegraphics[width=\textwidth]{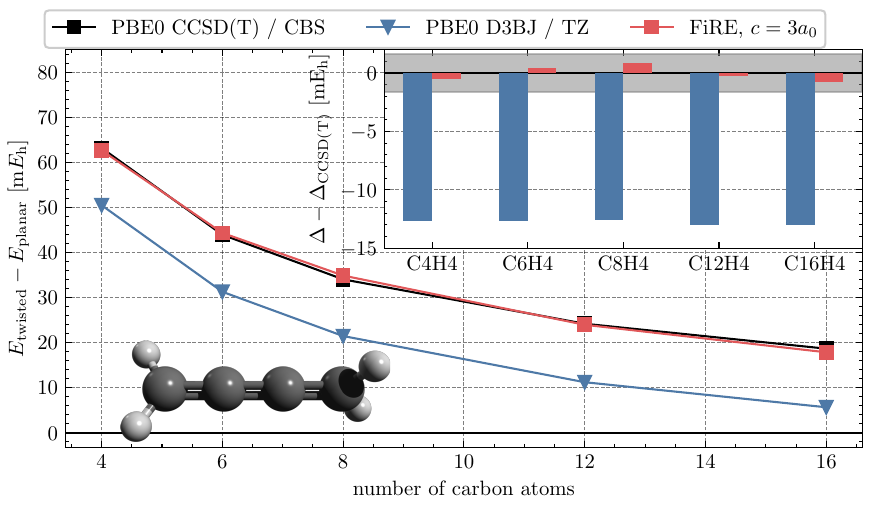}
    \end{minipage}
    
    \caption{\textbf{Cumulenes}: energy difference between twisted and planar geometry of cumulenes of increasing length, comparing our approach (FiRE) to CCSD(T) and density functional theory using the PBE0-functional. }
    \label{fig:cumulene_twist}
\end{figure*}

Cumulenes form an interesting test system because they contain long-range interactions.
For short chains, the equilibrium geometry is planar with a singlet ground state. 
The twisted geometry, with the methylene groups at each end twisted by 90 degrees, is higher in energy with a triplet ground state.
This system has been used to investigate long-range interactions in neural-network potentials \cite{frankSo3kratesEquivariantAttention2022}, and ethylene, the smallest of these molecules, has been used as a benchmark system for neural wave functions \cite{scherbelaTransferableFermionicNeural2024}.
We compute the energy difference between the twisted and planar geometry $E_\text{twisted} - E_\text{planar}$ for cumulenes of increasing size from $n=2$ to $n=16$ carbon atoms, using the $S^+$ spin operator \cite{liSpinsymmetryenforcedSolutionManybody2024} to enforce singlet and triplet states respectively.
\Cref{fig:cumulene_twist} depicts the energy difference as a function of the number of carbon atoms $n$, compared to several other quantum chemistry methods. 
We find that we can still accurately resolve this energy difference even with a small cutoff of $\cutoff=\qty{3}{\bohr}$, which is substantially smaller than the distance between the two methylene groups.
For short chain lengths where it is possible to run a CCSD(T) calculation, we find our method to be in good agreement with CCSD(T) with a maximum deviation of \qty{2}{mHa}.

Because we use this system as a simple benchmark system, we have not re-optimized the geometry for each geometry and spin state.
Thus, these energy differences may change when considering fully relaxed geometries.
Convergence of CCSD(T) calculations is nontrivial for this system due to strong spin contamination in unrestricted Hartree Fock calculations.
We find that only when using unrestricted Kohn Sham orbitals as a reference state --where spin contamination is much less severe -- does CCSD(T) converge to the correct solution.

\subsection{Model ablations}\label{sec:ablations}
\begin{figure*}[ht]
    \centering
    \includegraphics[width=\linewidth]{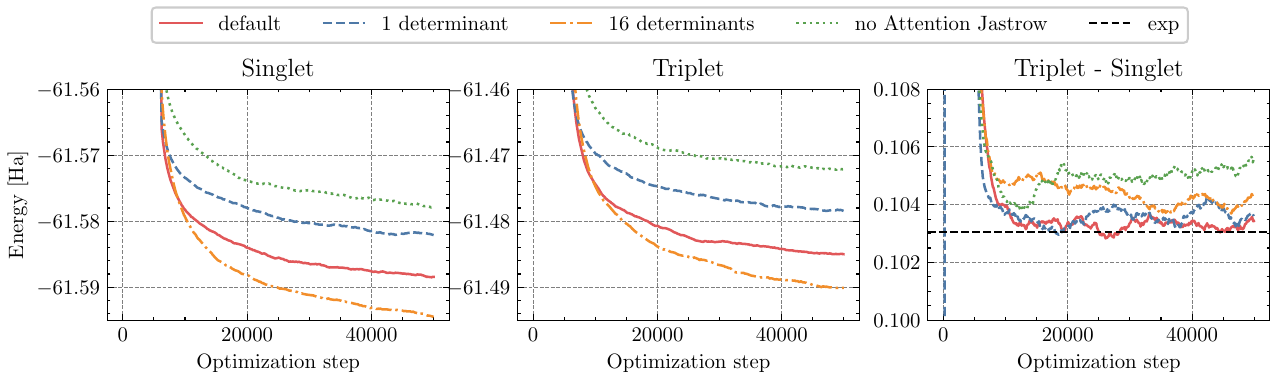}
    \caption{Ablation study on the network architecture.}
    \label{fig:ablation}
\end{figure*}

We investigate the importance of crucial hyperparameters of our neural wave function beyond the cutoff radius $\cutoff$.
For this, we investigate the singlet-triplet gap in naphthalene as in \cref{fig:relative_energies}c.
We compare four models, our standard FiRE with hyperparameters as defined in \cref{sec:hyperparameters}, one with only a single determinant $\ndet=1$, one with 16 determinants $\ndet=16$, and one without the attention Jastrow factor from \cref{sec:jastrow}.
The absolute energy for both states and the relative energy independence on the optimization steps are shown in \cref{fig:ablation}.
While enlarging the number of determinants to 16 improves absolute energies, convergence is slower, and the relative energy takes longer to converge.
Notably, FiRE accurately reconstructs the relative energy between the two states within 50k optimization steps, even with a single determinant.
Crucially, the attention Jastrow factor is important in accurately reconstructing the relative energy.

\subsection{Measuring speedups}
\label{sec:improved_scaling_empirical}
To test the effect of these speed-ups in practice, we compare the runtime of our ansatz against FermiNet \cite{pfauInitioSolutionManyelectron2020}, Psiformer \cite{vonglehnSelfAttentionAnsatzAbinitio2023} and LapNet \cite{liComputationalFrameworkNeural2024}.
For all three architectures we use the implementations in the LapNet codebase and use the forward Laplacian to accelerate kinetic energy computations~\citep{liComputationalFrameworkNeural2024}.
We use cumulenes, fully double-bonded hydrocarbon chains of the form \ch{CH_2=C_n=CH_2} as test systems.
We determine the runtime of all components required for a single optimization step: MCMC sampling, calculation of the kinetic energy, and potential evaluation of effective core potentials (ECP) and spin operators.
The two key runtimes are $\Tupd$, the time required to update the wave function after a single electron move (\cref{fig:scaling}a), and $\Tkin$, the time required to compute the kinetic energy (\cref{fig:scaling}b).
The total runtime $\Ttot$ per step is then given as
\begin{align}
    \Ttot &= \Tsamp + \Tkin + \Tecp + \Tspin\\
    \Tsamp &= \Twf + N_\text{sweeps}\, \nel\, \Tupd \\
    \Tecp &= \nel\, \Nquad\, \Tupd \\
    \Tspin &= \frac{\nel}{2}\, \Tupd
\end{align}
where in our experiments $N_\text{sweeps}=2$ is the number of Monte Carlo steps per electron, and $\Nquad=4$ is the number of quadrature points for estimating the non-local ECP.
For FermiNet, Psiformer, and LapNet, the time for a wave function update $\Tupd$ equals the time for a full wave function evaluation $\Twf$, whereas for our approach $\Tupd \ll \Twf$.
We use a batch size of 4096 samples on a single A100 GPU. 
For larger systems where not all samples fit into memory, we use the largest possible batch size per operation and method and scale the runtime accordingly.
To compare the empirical scaling of various methods, we fit the power laws of the form $T\sim\nel^\eta$.
We also compute energies for these cumulenes up to \ch{C_{16}H_4} and compare them to CCSD(T) in \cref{sec:cumulene}, finding good agreement.

\subsection{Low-rank updates in \texorpdfstring{$S_+$}{S+} operator}
To ensure pure states when comparing singlet and triplet states, we use the $S_+$ loss from \textcite{liSpinsymmetryenforcedSolutionManybody2024}.
There, in addition to minimizing the energy, we seek to minimize
\begin{align}
    P_{S_+} &= \left(\langle S_+ \wf\vert S_+\wf\rangle\right)^2\\
    \langle S_+ \wf\vert S_+\wf\rangle &= \frac{\Ndown}{\Nup + 1}\Ewf\left[R_{\beta}(\eposs)^2\right],\\
    R_{\beta}(\eposs)&=1 - \sum_{\alpha=0}^\Nup \frac{\wf(\pi_{\alpha,\beta}(\eposs))}{\wf(\eposs)}
\end{align}
where $\pi_{\alpha,\beta}$ is the permutation operator swapping the $\alpha$th electron with the $\beta$th electron.
Evaluating the wave function ratio involves evaluating the wave function with two electrons being permuted.
The gradient of the $P_{S_+}$ is given by 
\begin{align}
    \nabla_\theta P_{S_+} &= 2 P_+ \Ewf\left[2\left(R_\beta(\eposs)-P_+\right)\nabla_\theta\ln\wf(\eposs) + \nabla_\theta R_\beta(\eposs)\right]\label{eq:splus_grad}
\end{align}
Thanks to our local embeddings, we can efficiently compute this update to the wave function by only updating the electrons' embeddings within a $c$ radius of either swapped electron.
We efficiently compute the gradients of $R_\beta$ through our local updates in two parts.
Let $\vartheta$ denote the cached intermediate variables for our low-rank updates.
We decompose the gradient 
\begin{align}
    \nabla_\theta R_\beta(\rvec)=\frac{\partial R_\beta(\rvec)}{\partial \theta} + \frac{\partial R_\beta (\rvec)}{\partial \vartheta} \frac{\partial \vartheta}{\partial\theta}.
\end{align}
By aggregating $\frac{\partial R_\beta}{\partial \vartheta}$ for all swaps first, we avoid repeated backward passes for the gradient computation.

\subsection{Non-hermitian operator gradients in Spring}
We generally precondition gradients with Spring as in \cref{eq:spring}, though, this requires that the unpreconditioned gradient $\nabla_\theta\mathcal{L}$ of some loss $\mathcal{L}$ can be written as $\nabla_\theta\mathcal{L}=\mcqgt\frac{\partial \mathcal{L}}{\partial \ln\wf}$ like the energy gradient where $\frac{\partial E}{\partial \ln\wf}=\El(\rvec)-\Ewf\left[\El(\rvec)\right]$.
While any gradient of a hermitian operator can be written this way, it does not hold for non-hermitian operators like the $S_+$ operator due to the derivative through $R_\beta$ in \cref{eq:splus_grad}.
Thus, we would like to apply the general natural gradient update rule
\begin{align}
    \delta = \Ewf\left[\nabla_\theta\ln\wfp\nabla_\theta\ln\wfp^T\right]^{-1}\tilde{\delta}
\end{align}
for some general gradient $\tilde{\delta}$.
For a finite batch size, this can be written as
\begin{align}
    \delta = (\mcqgt\mcqgt^T)^{-1} \tilde{\delta}
\end{align}
which may be non-invertible if $\mcqgt\mcqgt^T$ is not full-rank.
Thus, one adds a damping factor to ensure invertibility 
\begin{align}
    \delta = (\mcqgt\mcqgt^T + \lambda I)^{-1} \tilde{\delta}.
\end{align}
which, after applying the Woodbury matrix identity, can be efficiently computed as 
\begin{align}
    \delta = \frac{1}{\lambda}\tilde{\delta} -  \mcqgt(\mcqgt^T\mcqgt + \lambda I)^{-1}\mcqgt^T \tilde{\delta}.
\end{align}
Crucially, if $\tilde{\delta}\notin\text{span}(\mcqgt)$, i.e., it cannot be written as $\tilde{\delta}=\mcqgt\tilde{\epsilon}$, the part that is not in $\mcqgt$ will be upscaled by $\frac{1}{\lambda}=1000$ for the typical choice of $\lambda=\frac{1}{1000}$.
This generally leads to unstable optimization.

We tackle this issue by splitting $\tilde{\delta}=\tilde{\delta}_\in + \tilde{\delta}_{\notin}$ into $\tilde{\delta}_\in\in\text{span}(\mcqgt)$ and $\tilde{\delta}_{\notin} =\tilde{\delta}-\tilde{\delta}_\in$, since we can write $\tilde{\delta}_\in=\mcqgt\tilde{\epsilon}$, we simply add it to $\epsilon$ in \cref{eq:spring}.
We add $\tilde{\delta}_{\notin}$ directly to the final gradient update.
Thus, the final gradient is 
\begin{align}
  \delta^t = \tilde{\delta}_{\notin} + \mcqgt \left(\mcqgt^T\mcqgt + \lambda I\right)^{-1}\left(\epsilon + \tilde{\epsilon} - \mcqgt\eta\delta^{t-1}\right) + \eta\delta^{t-1}.\label{eq:aux_spring}
\end{align}
To obtain the part that is within the span, we use the identity
\begin{align}
    \tilde{\delta}_\in=&\mcqgt(\mcqgt^T\mcqgt)^{-1}\mcqgt^T\tilde{\delta}=\mcqgt \mcqgt^+ \tilde{\delta}
\end{align}
where $\mcqgt^+$ is the Moore-Penrose pseudoinverse of $\mcqgt$, which we compute from the same hermitian eigendecomposition used to compute $(\mcqgt^T\mcqgt + \lambda I)^{-1}$.
Note that we compute $\tilde{\epsilon}=\mcqgt^+\tilde{\delta}$ in the process and use it for \cref{eq:aux_spring}.

\subsection{Effective core potential}
We use the cc-ECP by \textcite{bennettNewGenerationEffective2017}.
Unlike prior applications of ECPs to NN-VMC by \textcite{liFermionicNeuralNetwork2022}, we do not use a constant number of quadrature points $\Nquad$ to evaluate the non-local part but use a different $\Nquad$ per atom species.
For systems like chloroferrocene, with a single iron atom and 10 carbon atoms, we can substantially reduce the number of wave function evaluations by using $\Nquad=12$ for Fe but only $\Nquad=4$ for carbon, thus reducing the cost of ECP evaluation by $\approx 3\times$.
We also use effective core potentials for purely organic systems, such as acenes, where only 2 core electrons are removed per atom.
Due to the extra cost of evaluating the non-local ECP, we obtain little to no speed-up vs an all-electron calculation. However, we can substantially reduce the energy variance induced by the core electrons, thereby accelerating convergence.

\subsection{Energy extrapolation}
\label{sec:energy_extrapolation}
When computing interaction energies, the energies for both geometries do not necessarily converge at the same rate. 
Estimating the energy difference at a fixed number of optimization steps can, therefore, introduce a bias.
To reduce this effect's impact, we extrapolate each geometry's energy to its full-optimization limit.
\textcite{fuVarianceExtrapolationMethod2024} have proposed extrapolating the energy based on the energy variance, but we find that using the norm of the preconditioned energy gradients yields even better extrapolation accuracy.
Given iterates of the mean energy $E_t$ and gradient $\vg_t$ as a function of optimization steps $t$, we fit models of the form
\begin{equation}
    E_t = E_\infty + k |\vg_t|^2
\end{equation}
with the same slope $k$ for both geometries.
$E_\infty$ corresponds to the extrapolated energy, which would be obtained at the hypothetical limit of full convergence at zero gradients.
\Cref{tab:energies_s22} lists interaction energies with and without extrapolations, showing that extrapolation typically changes relative energies by less than \qty{1}{\mHa}, but removes a $\approx$~\qty{9}{\mHa} bias for H-bonded Uracil, where the dissociated geometry converges substantially faster. 
\begin{figure}[!ht]
    \centering
    \includegraphics[width=\textwidth]{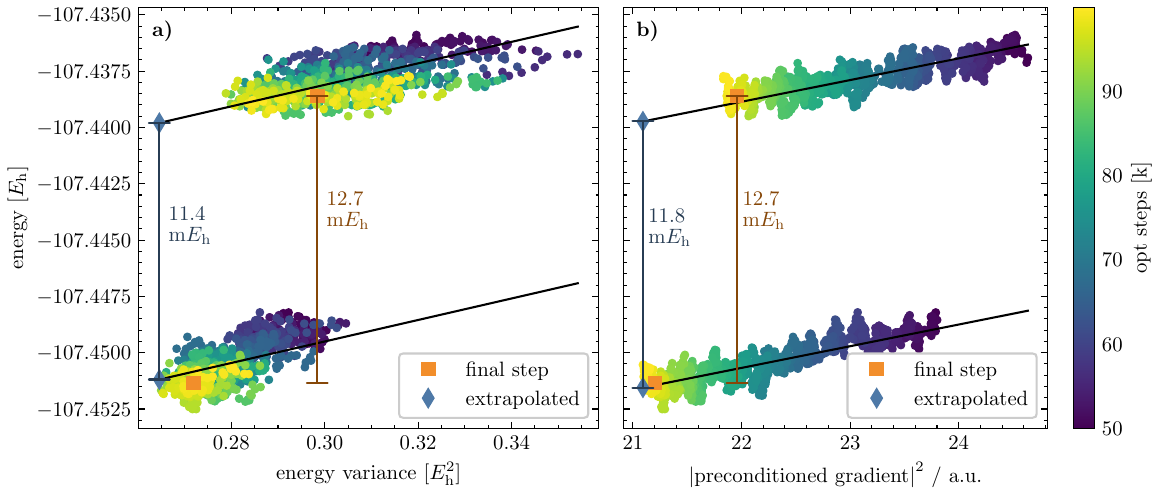}
    \caption{\textbf{Energy extrapolation for the phenol dimer} \textbf{a)} based on variance. \textbf{b)} based on preconditioned gradient norm. Energy as a function of variance / gradient norm for the dissociated geometry and the equilibrium geometry.}
    \label{fig:energy_extrapolation}
\end{figure}

\Cref{fig:energy_extrapolation} demonstrates the energy extrapolation on the example of the interaction energy of the phenol dimer. For this molecule, the equilibrium geometry converges slightly faster compared to the dissociated geometry, reaching lower energy, variance, and gradient norm for a given number of optimization steps.
Computing the energy difference after a fixed number of steps introduces a bias, which is remedied by extrapolating to the same variance or gradient norm.
\Cref{fig:energy_extrapolation} also shows that the gradient norm is less noisy and yields a better correlation with the energy compared to the variance.

\subsection{Implementation in JAX}
\label{sec:jax_implementation}
Like other neural-network VMC code~\citep{spencerBetterFasterFermionic2020}, we rely on JAX~\citep{bradburyJAXComposableTransformations2018} to accelerate our code on GPUs.
JAX traces the program to record tensor shapes and operations to create a directed acyclic graph (DAG) of the program.
This DAG is subsequently optimized and compiled into an accelerator-friendly program.
This process requires the tensor shapes to be fixed and known; calling the program with different input sizes triggers new time-intensive compiling processes.
On the one hand, using the largest possible tensor shapes eliminates the purpose of our finite-ranged embeddings and yields the same speed as running dense neural networks.
On the other hand, using tensor shapes that are too small results in incorrect computations, as non-zero elements need to be dropped.
To dynamically adjust to the current wave function and avoid an abundance of compilations, we compute the necessary tensor shapes to execute this step at every call.
In particular, we log the numbers of electron-nuclei edges, electron-electron edges, affected electrons in single-electron moves (local and global), electrons close to pseudopotentials, and the triplets in \cref{eq:logdet_lowrank}.
In subsequent steps, we use these as lower bounds with some padding as tensor shapes and pad the necessary tensors to fixed shapes, avoiding recompilation for every possible neighborhood combination.
This way, the first call may be numerically incorrect due to non-aligned tensor shapes, but subsequent calls minimize the amount of padding while maintaining exact computation.
Laplacian computations were done with \texttt{folx}~\citep{gaoFolxForwardLaplacian2023}.

\subsection{Theoretical VMC convergence rates}
\label{sec:theoretical_convergence_rates}
Recent work \cite{abrahamsenConvergenceVariationalMonte2024, liConvergenceAnalysisStochastic2024} has investigated theoretical convergence bounds for NN-VMC in conjunction with MCMC-based SGD methods. 
The theoretically established convergence rates are $\alpha = 1/4$ for convergence to a first-order stationary point \cite[Corollary 4.4]{abrahamsenConvergenceVariationalMonte2024} and $\alpha = 2/11$ for convergence to an approximate second-order stationary or a low-variance point \cite[Theorem 3]{liConvergenceAnalysisStochastic2024}.  
While the polynomial nature of these convergence results is consistent with our empirical findings, our empirical rate of $\alpha = 1$ is considerably higher than the rates suggested by theory. 
Potential reasons for this discrepancy could be our use of preconditioning during optimization or the fact that the variance of the sampling distribution tends to zero as an eigenvector is approached, thereby improving the sampling complexity. We also mention that under additional assumptions (most importantly a Polyak-Lojasiewicz condition), an optimal rate $\alpha = 1$ can be established for standard SGD-type methods \cite{khaledBetterTheorySGD2022}. 
While this rate would match our empirical findings, it is unclear if a  Polyak-Lojasiewicz condition holds in our setting. 
Furthermore, the results of \cite{abrahamsenConvergenceVariationalMonte2024,liConvergenceAnalysisStochastic2024} are not directly applicable to our setting because they monitor the loss gradient instead of the energy error and also rely on certain boundedness/mixing assumptions that may not be satisfied in our case. 
A more comprehensive analysis would be highly desirable but lies beyond the scope of this work.

\subsection{Conventional quantum chemistry calculations}
All conventional calculations were performed using ORCA 6.0.1 \cite{neeseSoftwareUpdateORCA2022} using correlation consistent basis cc-pVXZ sets by \textcite{dunningGaussianBasisSets1989} for CCSD(T) calculations and def2-XVP basis sets for DFT calculations.
We extrapolate results to the complete basis set (CBS) limit by extrapolating the Hartree-Fock energy from calculations at the triple- and quadruple-zeta levels.
The correlation energy is extrapolated from calculations performed at the double- and triple-zeta levels, where affordable.
Calculations denoted as \emph{DZ} correspond to Hartree-Fock energy at the CBS level and correlation energy obtained at the double-zeta level.
For basis set extrapolation, we use the relationships published by \textcite{neeseRevisitingAtomicNatural2011}.
For DFT calculations, we use the PBE0 \cite{perdewRationaleMixingExact1996} exchange-correlation-functional with the D3BJ dispersion correction \cite{beckeExchangeholeDipoleMoment2005}.
ZPVE for the benzene dimer at the MP2-level were taken from \cite{sinnokrotEstimatesInitioLimit2002}; for chloroferrocene, we calculated them at the PBE0/D3BJ level. 
Subtracting them from the experimental energies shifts the experimental relative energies by \qty{+0.55}{\mHa} and \qty{-0.2}{\mHa}, respectively.

\subsection{Hyperparameters}\label{sec:hyperparameters}
If not explicitly stated, experiments in this study use the hyperparameters provided in \cref{tab:hyperparameters}.
Notable hyperparameters are the cutoff $\cutoff$ that we investigate in \cref{sec:cutoff} and the number of optimization steps.
In general, one needs to increase the number of optimization steps with the system size.

\begin{table}
    \centering
    \caption{Hyperparameters}
    \label{tab:hyperparameters}
    \begin{tabular}{lc}
        \toprule
         Hyperparameter& Value \\
         \midrule
         Wave Function &\\
         \quad Determinants $\ndet$ & 4 \\
         \quad Cutoff $\cutoff$ & \\
         \quad\quad (non-)covalent interactions & \SI{5}{\bohr}\\
         \quad\quad ionization/singlet-triplet gap & \SI{3}{\bohr}\\
         \quad Cutoff $c_n$ & \SI{20}{\bohr}\\
         \quad Hidden dim $d$ & 256\\
         \quad Edge MLP widths & [16, 8]\\
         \quad Edge number of Gaussians $d_0$ & 32\\
         \quad Jastrow factor MLP widths & [256, 256] \\
         \quad Number of registers $\Nreg$ & 16\\
         \quad Register dimensions $\dreg$ & 16\\
         \quad Number of envelopes per nucleus & 8\\
         Pseudopotential &\\
         \quad ECP & ccECP\\
         \quad $\Nquad$, Li -- Ne & 4\\
         \quad $\Nquad$, Na -- Ar & 6\\
         \quad $\Nquad$, K -- Kr & 12\\
         Batch size $\Nwalker$ & 4096 \\
         Optimization &\\
         \quad Steps & 50.000\\
         \quad Learning rate & $\frac{0.1}{1 + \frac{t}{10000}}$\\
         \quad Damping $\lambda$ & 0.001\\
         \quad Spring decay $\eta$ & 0.99\\
         \quad Local energy clipping & 5 MAE\\
         \quad Clipping statistic & Median\\
         Spin operator gradient norm & 2\\
         MCMC &\\
         \quad Target acceptance ratio & \SI{50}{\percent}\\
         \quad Number of steps & $2\nel$\\
         \quad Number of global moves & 20\\
         Pretraining &\\
         \quad Basis set & ccecp-ccpvdz \\
         \quad Steps & 2000\\
         \quad Optimizer & Adam\\
         \quad Learning rate & $\frac{1}{1 + \frac{t}{1000}}$\\
         \bottomrule
    \end{tabular}
\end{table}

\subsection{Tables of energies}\label{sec:energies}
In the following, we list all energy estimates from \cref{fig:relative_energies}.
\cref{tab:energies_s22} lists the interaction energies for the S22 dataset, \cref{tab:benzene_dimer_energies} for the benzene dimer, \cref{tab:acene_energies} for the $n$-acenes, and \cref{tab:ferrocene_energies} for ferrocene.
\begin{table}
    \centering
    \caption{$n$-acene singlet-triplet gaps in mE$_\text{h}$, corresponding to \cref{fig:relative_energies}c}
    \label{tab:acene_energies}
\begin{tabular}{lrrrrr}
\toprule
 & \makecell[l]{Experiment\\(ZPE corrected)} & \makecell[l]{FiRE\\$c=3a_0$} & CCSD(T)/FPA & \makecell[l]{ACI-DSRG-\\MRPT2} & AFQMC \\
\midrule
naphthalene & 103.0 & 103.4(3) & 105.3 & 99.5 & 108.8(19) \\
anthracene & 72.6 & 74.4(4) & 77.1 & 69.1 & 73.9(19) \\
tetracene & 49.9 & 46.2(4) & 53.6 & 45.3 & 54.4(25) \\
pentacene & 34.1 & 34.5(5) & 40.5 & 28.8 & 40.3(25) \\
hexacene & 21.9 & 24.0(5) & 28.3 & 18.2 &  -- \\
\bottomrule
\end{tabular}
\end{table}

\begin{table}
    \centering
    \caption{Interaction energies for the S22 dataset in mE$_\text{h}$, corresponding to \cref{fig:relative_energies}a}
    \label{tab:energies_s22}
\begin{tabular}{l S[table-format=-2.2(1)] S[table-format=-2.2(1)] S[table-format=-2.2(1)] S[table-format=-2.2]}
\toprule
{molecule} & {\makecell[l]{FiRE, $c=5a_0$\\raw}} & {\makecell[l]{FiRE, $c=5a_0$\\extrapolated}} & {LapNet} & {CCSD(T)}\\
\midrule
Water dimer & 7.54(4) & 8.14(4) & 7.25(8) & 7.95\\
Formic acid dimer & 27.0(1) & 28.9(1) & 26.4(2) & 29.88\\
Formamide dimer & 23.2(1) & 24.4(1) & 22.6(1) & 25.60\\
Uracil dimer h-bonded & 21.7(2) & 31.7(2) & 37.0(4) & 32.89\\
Methane dimer & 0.69(3) & 0.91(3) & 0.3(1) & 0.84\\
Ethene dimer & 1.68(5) & 2.27(5) & 1.6(1) & 2.35\\
Uracil dimer stack & 11.0(2) & 14.7(2) & 11.7(4) & 15.63\\
Ethene-ethyne complex & 2.06(5) & 2.43(5) & 1.30(8) & 2.38\\
Benzene-water complex & 4.77(9) & 5.16(9) & 2.9(2) & 5.22\\
Benzene dimer T-shaped & 3.4(1) & 3.6(1) & 2.6(1) & 4.33\\
Phenol dimer & 12.7(2) & 11.8(2) & 8.0(3) & 11.31\\
\bottomrule
\end{tabular}
\end{table}

\begin{table}
    \begin{minipage}[t]{0.48\textwidth}
        \centering
        \caption{Benzene dimer binding energy in mE$_\text{h}$, corresponding to \cref{fig:relative_energies}b}
        \label{tab:benzene_dimer_energies}
        \begin{tabular}{l S[table-format=-2.1(1)]}
            \toprule
{method} & {interaction energy}\\
\midrule
Experiment, Grover et al & 4.4(8)\\
Experiment, Krause et al & 3.1(4)\\
CCSD(T)/CBS, Marshall et al & 4.3\\
FermiNet VMC, Ren et al & 18.2(6)\\
FermiNet DMC, Ren et al & 9.2(5)\\
FermiNet VMC, Glehn et al & 4.6(8)\\
Psiformer, Glehn et al & 0.7(3)\\
LapNet, Li et al & 2.6(1)\\
FiRE, $c=3a_0$, raw & 2.3(2)\\
FiRE, $c=3a_0$, extrapolated & 2.9(2)\\
FiRE, $c=5a_0$, raw & 3.4(1)\\
FiRE, $c=5a_0$, extrapolated & 3.6(1)\\
FiRE, $c=7a_0$, raw & 4.1(1)\\
FiRE, $c=7a_0$, extrapolated & 4.6(1)\\
            \bottomrule
        \end{tabular}
    \end{minipage}%
    \hfill
    \begin{minipage}[t]{0.48\textwidth}
        \centering
        \caption{Energy difference between (singlet) and twisted (triplet) cumulene in mE$_\text{h}$, corresponding to \cref{fig:cumulene_twist}}
        \label{tab:cumulene_energies}
        \begin{tabular}{l S[table-format=-2.2(1)] S[table-format=-2.2(1)] S[table-format=-2.2]}
            \toprule
            {molecule} & {FiRE, $c=3a_0$} & {CCSD(T)} & {PBE0}\\
            \midrule
            \ch{C_{4}H4} & 62.7(1) & 63.2 & 50.5 \\
            \ch{C_{6}H4} & 44.3(2) & 43.9 & 31.3 \\
            \ch{C_{8}H4} & 34.9(2) & 34.0 & 21.5 \\
            \ch{C_{12}H4} & 24.0(3) & 24.2 & 11.2 \\
            \ch{C_{16}H4} & 17.9(3) & 18.7 & 5.7 \\
            \bottomrule
        \end{tabular}
    \end{minipage}
\end{table}

\begin{table}
        \centering
        \caption{Chloroferrocene ionization potential in mE$_\text{h}$, corresponding to \cref{fig:organometallic}a}
        \label{tab:ferrocene_energies}
        \begin{tabular}{l S[table-format=3.1(1)]}
            \toprule
{method} & {IP}\\
\midrule
Experiment & 257.8\\
B3LYP/TZ & 237.4\\
PBE0/CBS & 242.5\\
DLPNO-CCSD(T)/CBS & 249.6\\
CCSD(T)/DZ & 245.3\\
CCSD(T)/FPA & 255.8\\
FiRE & 256.1(3)\\
            \bottomrule
        \end{tabular}
\end{table}

\begin{table}
    \centering
        \caption{Energies corresponding to \cref{fig:organometallic}b. Relative energies for the four protonation sites and mean absolute error (MAE) to the conventional best estimate, in mE$_\text{h}$}
        \label{tab:Fe2S2_energies}
        \begin{tabular}{l S[table-format=-2.1(1)] S[table-format=-2.1(1)] S[table-format=-2.1(1)] S[table-format=-2.1(1)] | S[table-format=2.1(1)]}
            \toprule
{method} & {HC} & {HS} & {HFe} & {HFe2} & {MAE}\\
\midrule
Conventional best est. & -53.9 &-2.3 &33.9 &22.2 &0.0\\
\midrule
FiRE, $c=5a_0$, raw & -56.2(3) &-0.1(3) &33.7(3) &22.6(3) &1.3(2)\\
No (T): CCSD/CBS+DMRG & -67.0 &-11.4 &41.0 &37.4 &11.1\\
No CBS: CCSD(T)/TZ+DMRG & -49.2 &1.6 &28.7 &18.9 &4.3\\
No DMRG: CCSD(T)/CBS & -57.3 &-3.7 &34.9 &26.1 &2.4\\
            \bottomrule
        \end{tabular}
\label{tab:energies_Fe2S2}
\end{table}

\clearpage
\printbibliography[title={Supplementary Information References}]
\end{document}